\documentclass[preprint,prd,aps,showpacs,showkeys,nofootinbib]{revtex4}
\usepackage{graphicx}
\usepackage{dcolumn}
\usepackage{bm}
\begin{document}

\title{Scalar neutrino dark matter in $U(1)_X$SSM}
\author{Shu-Min Zhao$^{1,2}$\footnote{zhaosm@hbu.edu.cn}, Tai-Fu Feng$^{1,2,3}$\footnote{fengtf@hbu.edu.cn},
Ming-Jie Zhang$^{1,2}$, Jin-Lei Yang$^{1,2}$, Hai-Bin Zhang$^{1,2}$, Guo-Zhu Ning$^{1,2}$}

\affiliation{$^1$ Department of Physics, Hebei University, Baoding 071002, China}
\affiliation{$^2$ Key Laboratory of High-precision Computation and Application of Quantum Field Theory of Hebei Province, Baoding 071002, China}
\affiliation{$^3$ Department of Physics, Chongqing University, Chongqing 401331, China}
\date{\today}

\begin{abstract}

 $U(1)_X$SSM is the extension of the minimal supersymmetric standard model(MSSM) and its local
 gauge group is $SU(3)_C\times SU(2)_L \times U(1)_Y \times U(1)_X$. To obtain this model, three singlet new Higgs superfields and right-handed neutrinos are added to MSSM.
In the framework of $U(1)_X$SSM, we study the Higgs mass and take
 the lightest CP-even sneutrino as a cold dark matter candidate.
For the lightest CP-even sneutrino, the relic density and the cross section for dark matter scattering off nucleon are both researched.
In suitable parameter space of the model, the numerical results satisfy the constraints of the relic density and
the cross section with the nucleon.

\end{abstract}


\maketitle

\section{introduction}

From the cosmological observations, astronomers are sure about the existence of dark matter in the universe,
 whose contribution is about five times that of visible matter \cite{account1,account2}. Various luminous objects (stars,
 gas clouds globular clusters, or entire galaxies), moving faster than  expectations \cite{rotation1,rotation2}, are the
 earliest and the most compelling evidences for dark matter \cite{other exist1,other exist2,other exist3,other exist4}.
Dark matter must be electrically and color neutral, and can
only take part in weak interactions.
Dark matter is stable and has a long life-time \cite{longlife1,longlife2}.
At present, the mass and interaction properties of the dark matter are unknown.

 Though the standard model(SM) successfully predicts the detection of the CP-even Higgs(125.1GeV) \cite{mh01,mh02},
  it can not explain the relic density of dark matter in the universe.
  The relic density of light neutrinos with tiny mass is
 $\Omega_\nu h^2\leq0.0062$ at 95\% confidence level, that is much smaller than non-baryonic matter density
 $\Omega_\nu h^2=0.1186\pm0.0020$ \cite{pdg}. As a result, there must exist new physics beyond the SM.
 There are several dark matter candidates: axions, sterile neutrinos, primordial black holes and weakly interacting massive
 particles (WIMPs) \cite{longlife1,WIMP,WIMP1}. WIMP, in particular, ranks among the most popular candidates for dark matter, whose detection is crucial for both
 distinguishing new physics models and understanding the nature of dark matter.  The direct detection for dark matter is
 studying the recoil energy of nuclei caused by the elastic scattering of a WIMP off a nucleon.

   The neutralino in the minimal supersymmetric standard model (MSSM) has been extensively studied \cite{MSSM} as one of the favorite dark matter candidates.
   However, the left-handed sneutrino meets severe troubles because the cross section for elastic scattering off nuclei exceeds
the experimental limit by several orders with the exchange of vector boson Z \cite{LSneu}. Considering the neutrino oscillations, neutrino
should possess tiny mass \cite{neutrino1,neutrino2}. Thus, to obtain light neutrino mass, one can add right-handed neutrino to the MSSM. The supersymmetric partners of the right-handed neutrinos will
provide an alternative dark matter candidate \cite{Sneudark11,Sneudark12,Sneudark13,Sneudark14,Sneudark15,Caojunjie1,Caojunjie2}.
There are also other works on sneutrino dark matter \cite{Sneudark13, TaoHan, Sneudark21, Sneudark22, Sneudark23, Sneudark24,  sneutrinoD1,sneutrinoD2,sneutrinoD3}.
At last but not least, it is worth mentioning that U(1) extensions of the MSSM considered in the works \cite{UMSSM1,UMSSM2,UMSSM3,UMSSM4,UMSSM5} have been of great interest lately.

In this work, we extend the MSSM to the $U(1)_X$SSM, whose local
gauge group is $SU(3)_C\times SU(2)_L \times U(1)_Y\times U(1)_X$ \cite{Sarah1,Sarah2,Sarah3}.
In comparison with the MSSM, our model $U(1)_X$SSM has more superfields: $U(1)_X$ gauge field,
righ-handed neutrinos, three $SU(2)_L$ singlet Higgs superfields $\hat{\eta},~\hat{\bar{\eta}},~\hat{S}$ and their superpartners.
The vacuum expectation
value(VEV) of $\bar{\eta}$ produces masses of the right-handed neutrinos.
The righ-handed neutrinos and left-handed neutrinos mix together through $Y_\nu\hat{\nu}\hat{l}\hat{H}_u$. Therefore, light neutrinos
obtain  tiny masses through the seesaw mechanism. The lightest sneutrino can be a new dark matter candidate different from the case in MSSM.
Moreover, PAMELA \cite{PAMELA} claims an excess in the electron/positron flux and no excess in the proton/antiproton flux \cite{pzhandpfu}. Thus, the idea that
dark matter carries lepton number is intriguing.
The little hierarchy problem in MSSM is relieved in $U(1)_X$SSM by the
right-handed neutrinos, sneutrinos and additional Higgs singlets. $U(1)_X$SSM includes both terms $\mu\hat{H}_u\hat{H}_d$ and
$\lambda_H\hat{S}\hat{H}_u\hat{H}_d$. When $\hat{S}$ develops a VEV ($v_S/\sqrt{2}$), an effective $\mu_{eff}$ is obtained as
$\mu_{eff}=\mu+\lambda_Hv_S/\sqrt{2}$, which can relieve the $\mu$ problem and even solve it.
The spontaneously broken $U(1)_X$ gauge symmetry can be used to avoid baryon number violating operators and keep proton stable.
The interaction between three extra singlet Higgs superfields and two Higgs doublets is favorable to increase the mass of the lightest
CP-even Higgs at the tree level. At the same time, the $U(1)_X$ D-term gives another contribution. Considering  both effects, large
loop-induced contribution from stop sector is not necessary. Furthermore, the mass of the next light CP-even Higgs can reach the order of TeV.
The added parameters mitigate the constraints from experiments such as LHC.

We introduce the $U(1)_X$SSM in detail in section II.
Supposing the lightest CP-even sneutrino as a dark matter candidate, we study its relic density in section III. Section IV is devoted to research the
direct detection for sneutrino elastic scattering off the nuclei. The numerical results for Higgs masses,
relic density for dark matter and its direct detection
are all presented in section V. Sec. VI is devoted to the discussions and conclusions.

\section{the $U(1)_X$SSM}

The  gauge group of the $U(1)_X$SSM is $SU(3)_C\otimes
SU(2)_L \otimes U(1)_Y\otimes U(1)_X$.  To obtain the $U(1)_X$SSM,  new superfields are added to the MSSM, namely:
three Higgs singlets $\hat{\eta},~\hat{\bar{\eta}},~\hat{S}$ and right-handed neutrinos $\hat{\nu}_i$. It can give light
neutrino mass at the tree level through the seesaw mechanism. The neutral CP-even parts of
$H_u,~ H_d,~\eta,~\bar{\eta}$ and $S$ mix together, forming $5\times5 $ mass squared matrix.
The loop corrections to the lightest CP-even Higgs are important and they are taken into account to get 125 GeV Higgs mass \cite{LCTHiggs1,LCTHiggs2}.
\begin{table}
\caption{ The superfields in $U(1)_X$SSM}
\begin{tabular}{|c|c|c|c|c|}
\hline
Superfields & $SU(3)_C$ & $SU(2)_L$ & $U(1)_Y$ & $U(1)_X$ \\
\hline
$\hat{Q}_i$ & 3 & 2 & 1/6 & 0 \\
\hline
$\hat{u}^c_i$ & $\bar{3}$ & 1 & -2/3 & -$1/2$ \\
\hline
$\hat{d}^c_i$ & $\bar{3}$ & 1 & 1/3 & $1/2$  \\
\hline
$\hat{L}_i$ & 1 & 2 & -1/2 & 0  \\
\hline
$\hat{e}^c_i$ & 1 & 1 & 1 & $1/2$  \\
\hline
$\hat{\nu}_i$ & 1 & 1 & 0 & -$1/2$ \\
\hline
$\hat{H}_u$ & 1 & 2 & 1/2 & 1/2\\
\hline
$\hat{H}_d$ & 1 & 2 & -1/2 & -1/2 \\
\hline
$\hat{\eta}$ & 1 & 1 & 0 & -1 \\
\hline
$\hat{\bar{\eta}}$ & 1 & 1 & 0 & 1\\
\hline
$\hat{S}$ & 1 & 1 & 0 & 0 \\
\hline
\end{tabular}
\label{quarks}
\end{table}

The superpotential for this model reads:
\begin{eqnarray}
&&W=l_W\hat{S}+\mu\hat{H}_u\hat{H}_d+M_S\hat{S}\hat{S}-Y_d\hat{d}\hat{q}\hat{H}_d-Y_e\hat{e}\hat{l}\hat{H}_d+\lambda_H\hat{S}\hat{H}_u\hat{H}_d
\nonumber\\&&+\lambda_C\hat{S}\hat{\eta}\hat{\bar{\eta}}+\frac{\kappa}{3}\hat{S}\hat{S}\hat{S}+Y_u\hat{u}\hat{q}\hat{H}_u+Y_X\hat{\nu}\hat{\bar{\eta}}\hat{\nu}
+Y_\nu\hat{\nu}\hat{l}\hat{H}_u.
\end{eqnarray}

There are two Higgs doublets and three Higgs singlets, whose explicit forms are shown in the follow,
\begin{eqnarray}
&&H_{u}=\left(\begin{array}{c}H_{u}^+\\{1\over\sqrt{2}}\Big(v_{u}+H_{u}^0+iP_{u}^0\Big)\end{array}\right),
~~~~~~
H_{d}=\left(\begin{array}{c}{1\over\sqrt{2}}\Big(v_{d}+H_{d}^0+iP_{d}^0\Big)\\H_{d}^-\end{array}\right),
\nonumber\\
&&\eta={1\over\sqrt{2}}\Big(v_{\eta}+\phi_{\eta}^0+iP_{\eta}^0\Big),~~~~~~~~~~~~~~~
\bar{\eta}={1\over\sqrt{2}}\Big(v_{\bar{\eta}}+\phi_{\bar{\eta}}^0+iP_{\bar{\eta}}^0\Big),\nonumber\\&&
\hspace{4.0cm}S={1\over\sqrt{2}}\Big(v_{S}+\phi_{S}^0+iP_{S}^0\Big).
\end{eqnarray}
$v_u,~v_d,~v_\eta$,~ $v_{\bar\eta}$ and $v_S$ are the corresponding  VEVs of the Higgs superfields $H_u$, $H_d$, $\eta$, $\bar{\eta}$ and $S$.
Here, we define $\tan\beta=v_u/v_d$ and $\tan\beta_\eta=v_{\bar{\eta}}/v_{\eta}$. The definition of
$\tilde{\nu}_L$ and $\tilde{\nu}_R$ is
\begin{eqnarray}
\tilde{\nu}_L=\frac{1}{\sqrt{2}}\phi_l+\frac{i}{\sqrt{2}}\sigma_l,~~~~~~~~~~\tilde{\nu}_R=\frac{1}{\sqrt{2}}\phi_R+\frac{i}{\sqrt{2}}\sigma_R.
\end{eqnarray}

The soft SUSY breaking terms are
\begin{eqnarray}
&&\mathcal{L}_{soft}=\mathcal{L}_{soft}^{MSSM}-B_SS^2-L_SS-\frac{T_\kappa}{3}S^3-T_{\lambda_C}S\eta\bar{\eta}
+\epsilon_{ij}T_{\lambda_H}SH_d^iH_u^j\nonumber\\&&
-T_X^{IJ}\bar{\eta}\tilde{\nu}_R^{*I}\tilde{\nu}_R^{*J}
+\epsilon_{ij}T^{IJ}_{\nu}H_u^i\tilde{\nu}_R^{I*}\tilde{l}_j^J
-m_{\eta}^2|\eta|^2-m_{\bar{\eta}}^2|\bar{\eta}|^2\nonumber\\&&
-m_S^2S^2-(m_{\tilde{\nu}_R}^2)^{IJ}\tilde{\nu}_R^{I*}\tilde{\nu}_R^{J}
-\frac{1}{2}\Big(M_X\lambda^2_{\tilde{X}}+2M_{BB^\prime}\lambda_{\tilde{B}}\lambda_{\tilde{X}}\Big)+h.c~~.
\end{eqnarray}

The particle content and charge assignments for $U(1)_X$SSM are shown in the Table 1.
 We use $Y^Y$ for representing the $U(1)_Y$ charge and $Y^X$ for representing the $U(1)_X$ charge.
According to the textbook \cite{Peskin}, the SM is
anomaly free. The details regarding the absence of anomaly within the $U(1)_X$SSM model can be summarized as follows:

1. The anomaly of three $SU(2)_L$ gauge bosons vanishes as in the SM and the condition of three $SU(3)_C$ gauge bosons is similar.

2. The anomalies containing one $SU(3)_C$ boson or one $SU(2)_L$ boson are proportional to $Tr[t^a]=0$ or $Tr[\tau^a]=0$.

3.  The anomaly of one $U(1)_Y$ or  $U(1)_X$ boson with two $SU(3)_C$ bosons is proportional to the group theory factor
$Tr[t^at^bY^Y]=\frac{1}{2}\delta^{ab}\sum_q Y^Y_q$ or $Tr[t^at^bY^X]=\frac{1}{2}\delta^{ab}\sum_q Y^X_q$.

4. The anomaly of one $U(1)_Y$ or  $U(1)_X$ boson with two $SU(2)_L$ bosons is proportional to
$Tr[\tau^a\tau^bY^Y]=\frac{1}{2}\delta^{ab}\sum_{L} Y^Y_{L}$ or $Tr[\tau^a\tau^bY^X]=\frac{1}{2}\delta^{ab}\sum_{L} Y^X_{L}$.

5. The anomalies of three U(1) gauge bosons are divided into four types
\begin{eqnarray}
&&Tr[Y^YY^YY^Y]=\sum_n(Y^Y_n)^3,~~~~~~~~~~~Tr[Y^XY^XY^X]=\sum_n(Y^X_n)^3,\nonumber\\&&
Tr[Y^XY^YY^Y]=\sum_nY^X_n(Y^Y_n)^2,~~~~~~Tr[Y^YY^XY^X]=\sum_nY^Y_n(Y^X_n)^2.
\end{eqnarray}

6. The gravitational anomaly with one U(1) gauge boson is proportional to $Tr[Y^Y]=\sum_n Y_n^Y$ or $Tr[Y^X]=\sum_n Y_n^X$.

The anomalies that do not relate to $U(1)_X$  are very similar as the SM condition and can be proved free easily.
The anomalies including $U(1)_X$ are also proved free, which are more complicated than those of SM.
In the end, this model is anomaly free.

The presence of two Abelian groups $U(1)_Y$ and $U(1)_X$ in $U(1)_X$SSM has a new effect absent in the MSSM with just one Abelian gauge group $U(1)_Y$:
the gauge kinetic mixing. This effect can also be induced through RGEs, even if it is set to zero at $M_{GUT}$.

The covariant derivatives of this model have the general form \cite{UMSSM5,B-L1,B-L2,gaugemass}
\begin{eqnarray}
&&D_\mu=\partial_\mu-i\left(\begin{array}{cc}Y,&X\end{array}\right)
\left(\begin{array}{cc}g_{Y},&g{'}_{{YX}}\\g{'}_{{XY}},&g{'}_{{X}}\end{array}\right)
\left(\begin{array}{c}A_{\mu}^{\prime Y} \\ A_{\mu}^{\prime X}\end{array}\right)\;.
\label{gauge1}
\end{eqnarray}
 Here, $A_{\mu}^{\prime Y}$ and $A^{\prime X}_\mu$ denote the gauge fields of $U(1)_Y$ and $U(1)_X$,
  while $Y$ and $X$ represent the hypercharge and $X$ charge respectively. We can perform a basis transformation,
  because the two Abelian gauge groups are unbroken.
  The following formula can be obtained with a correct matrix $R$ \cite{UMSSM5,B-L2,gaugemass}
  \begin{eqnarray}
&&\left(\begin{array}{cc}g_{Y},&g{'}_{{YX}}\\g{'}_{{XY}},&g{'}_{{X}}\end{array}\right)
R^T=\left(\begin{array}{cc}g_{1},&g_{{YX}}\\0,&g_{{X}}\end{array}\right)\;.
\label{gauge3}
\end{eqnarray}

So the $U(1)$ gauge fields are redefined as
\begin{eqnarray}
&&R\left(\begin{array}{c}A_{\mu}^{\prime Y} \\ A_{\mu}^{\prime X}\end{array}\right)
=\left(\begin{array}{c}A_{\mu}^{Y} \\ A_{\mu}^{X}\end{array}\right)\;.
\label{gauge4}
\end{eqnarray}

The interesting thing is that the gauge bosons $A^{X}_\mu,~A^Y_\mu$ and $V^3_\mu$ mix together at the tree level, and the mass matrix
is shown in the basis $(A^Y_\mu, V^3_\mu, A^{X}_\mu)$
\begin{eqnarray}
&&\left(\begin{array}{*{20}{c}}
\frac{1}{8}g_{1}^2 v^2 &~~~ -\frac{1}{8}g_{1}g_{2} v^2 & ~~~\frac{1}{8}g_{1}g_{{YX}} v^2 \\
-\frac{1}{8}g_{1}g_{2} v^2 &~~~ \frac{1}{8}g_{2}^2 v^2 & ~~~~-\frac{1}{8}g_{2}g_{{YX}} v^2\\
\frac{1}{8}g_{1}g_{{YX}} v^2 &~~~ -\frac{1}{8}g_{2}g_{{YX}} v^2 &~~~~ \frac{1}{8}g_{{YX}}^2 v^2+\frac{1}{8}g_{{X}}^2 \xi^2
\end{array}\right),\label{gauge matrix}
\end{eqnarray}
with $v^2=v_u^2+v_d^2$ and $\xi^2=v_\eta^2+v_{\bar{\eta}}^2$.
To diagonalize the mass matrix in Eq. (\ref{gauge matrix}), an unitary matrix
including two mixing angles $\theta_{W}$ and $\theta_{W}'$ is used here
\begin{eqnarray}
&&\left(\begin{array}{*{20}{c}}
\gamma_\mu\\ [6pt]
Z_\mu\\ [6pt]
Z'_\mu
\end{array}\right)=
\left(\begin{array}{*{20}{c}}
\cos\theta_{W} & \sin\theta_{W} & 0 \\ [6pt]
-\sin\theta_{W}\cos\theta_{W}' & \cos\theta_{W}\cos\theta_{W}' & \sin\theta_{W}'\\ [6pt]
\sin\theta_{W}\sin\theta_{W}' & -\cos\theta_{W}'\sin\theta_{W}' & \cos\theta_{W}'
\end{array}\right)
\left(\begin{array}{*{20}{c}}
A^Y_\mu\\ [6pt]
V^3_\mu\\ [6pt]
A^{X}_\mu
\end{array}\right).
\end{eqnarray}
We deduce $\sin^2\theta_{W}^\prime$ as
\begin{eqnarray}
\sin^2\theta_{W}'=\frac{1}{2}-\frac{(g_{{YX}}^2-g_{1}^2-g_{2}^2)v^2+
4g_{X}^2\xi^2}{2\sqrt{(g_{{YX}}^2+g_{1}^2+g_{2}^2)^2v^4+8g_{X}^2(g_{{YX}}^2-g_{1}^2-g_{2}^2)v^2\xi^2+16g_{X}^4\xi^4}}.
\end{eqnarray}
The new mixing angle $\theta_{W}^\prime$ appears in the couplings involving $Z$ and $Z^{\prime}$. The exact eigenvalues of Eq. (\ref{gauge matrix})
are calculated \cite{UMSSM5,B-L2,gaugemass}
\begin{eqnarray}
&&\qquad\;\quad\;m_\gamma^2=0,\nonumber\\
&&\qquad\;\quad\;m_{Z,{Z^{'}}}^2=\frac{1}{8}\Big((g_{1}^2+g_2^2+g_{YX}^2)v^2+4g_{X}^2\xi^2 \nonumber\\
&&\qquad\;\qquad\;\qquad\;\mp\sqrt{(g_{1}^2+g_{2}^2+g_{YX}^2)^2v^4+8(g_{YX}^2-g_{1}^2-
g_{2}^2)g_{X}^2v^2\xi^2+16g_{X}^4\xi^4}\Big).
\end{eqnarray}

The Higgs potential is deduced here
\begin{eqnarray}
&&V=\frac{1}{2}g_X(g_X+g_{YX})(|H_d^0|^2-|H_u^0|^2)(|\eta|^2-|\bar{\eta}|^2)+|\lambda_H|^2|H_u^0H_d^0|^2+m^2_{S}|S|^2\nonumber\\&&
+\frac{1}{8}\Big(g_1^2+g_2^2+(g_X+g_{YX})^2\Big)(|H_d^0|^2-|H_u^0|^2)^2+\frac{1}{2}g_X^2(|\eta|^2-|\bar{\eta}|^2)^2+\lambda_C^2|\eta\bar{\eta}|^2
\nonumber\\&&
+(|\mu|^2+|\lambda_H|^2|S|^2+2\mathrm{Re}[\mu^*\lambda_HS])(|H_d^0|^2+|H_u^0|^2)+|\lambda_C|^2|S|^2(|\eta|^2+|\bar{\eta}|^2)
\nonumber\\&&+2\mathrm{Re}[l_W^*(2M_SS+\lambda_C\eta\bar{\eta}-\lambda_HH_u^0H_d^0
+\kappa S^2)]+4|M_S|^2|S|^2+
2\mathrm{Re}[\lambda_C^*\kappa\eta^*\bar{\eta}^*S^2]
\nonumber\\&&+|\kappa|^2|S|^4+4\mathrm{Re}[M_S^*S^*(\lambda_C\eta\bar{\eta}-\lambda_HH_u^0H_d^0+\kappa S^2)]
-2\mathrm{Re}[\lambda_C^*\lambda_H\eta^*\bar{\eta}^*H_u^0H_d^0]+|l_W|^2\nonumber\\&&-2\mathrm{Re}[B_\mu H_d^0H_u^0]+ 2\mathrm{Re}[L_S S]
+\frac{2}{3}\mathrm{Re}[T_kS^3]+ 2\mathrm{Re}[T_{\lambda_C}\eta\bar{\eta}S]-2\mathrm{Re}[T_{\lambda_H}H_d^0H_u^0 S]\nonumber\\&&-2\mathrm{Re}[\lambda_H\kappa^* H_u^0H_d^0 (S^2)^*]
+m^2_\eta|\eta|^2+m^2_{\bar{\eta}}|\bar{\eta}|^2+
m^2_{H_u^0}|H_u^0|^2+m^2_{H_d^0}|H_d^0|^2+2\mathrm{Re}[B_S S^2].\label{V}
\end{eqnarray}

 To simplify the following discussion, we suppose that the parameters
($\mu,~\lambda_H,~\lambda_C,~l_W,~ M_S,~B_\mu,~L_S,~T_{\kappa},~ T_{\lambda_C},~ T_{\lambda_H},~\kappa,~B_S$) in Eq. (\ref{V}) are real parameters.
The VEVs of the Higgs satisfy the following equations
\begin{eqnarray}
&&\frac{1}{8}\Big(g_1^2+g_2^2+(g_X+g_{YX})^2\Big)(v_d^2-v_u^2)v_d+
\frac{1}{4}g_X(g_X+g_{YX})v_d(v^2_{\eta}-v^2_{\bar{\eta}})\nonumber\\&&+
(\mu^2+\frac{1}{2}\lambda_H^2v_S^2+\sqrt{2}\mu\lambda_H v_S)v_d-l_W\lambda_Hv_u+\frac{1}{2}\lambda_H^2v_u^2v_d
-\sqrt{2}M_S\lambda_Hv_Sv_u\nonumber\\&&-\frac{1}{2}\lambda_H\lambda_Cv_{\eta}v_{\bar{\eta}}v_u
-\frac{1}{2}\lambda_H\kappa v_u v_S^2+m_{H_d}^2v_d
-B_\mu v_u-\frac{T_{\lambda_H}}{\sqrt{2}}v_uv_S=0,\\&&
\frac{1}{8}\Big(g_1^2+g_2^2+(g_X+g_{YX})^2\Big)(v_u^2-v_d^2)v_u+
\frac{1}{4}g_X(g_X+g_{YX})v_u(v^2_{\bar{\eta}}-v^2_{\eta})\nonumber\\&&+
(\mu^2+\frac{1}{2}\lambda_H^2v_S^2+\sqrt{2}\mu\lambda_H v_S)v_u-l_W\lambda_Hv_d+\frac{1}{2}\lambda_H^2v_uv_d^2
-\sqrt{2}M_S\lambda_Hv_Sv_d\nonumber\\&&-\frac{1}{2}\lambda_H\lambda_Cv_{\eta}v_{\bar{\eta}}v_d
-\frac{1}{2}\lambda_H\kappa v_d v_S^2+m_{H_u}^2v_u
-B_\mu v_d-\frac{T_{\lambda_H}}{\sqrt{2}}v_dv_S=0,
\\&&\frac{1}{2}g_X^2(v_{\eta}^2-v_{\bar{\eta}}^2)v_\eta-
\frac{1}{4}g_X(g_X+g_{YX})v_\eta(v^2_{u}-v^2_{d})
+\frac{1}{2}\lambda_C^2v_\eta (v_S^2+v_{\bar{\eta}}^2)+l_W\lambda_Cv_{\bar{\eta}}\nonumber\\&&
+\sqrt{2}M_S\lambda_Cv_Sv_{\bar{\eta}}-\frac{1}{2}\lambda_H\lambda_Cv_{\bar{\eta}}v_{u}v_d
+\frac{1}{2}\lambda_C\kappa v_{\bar{\eta}} v_S^2+m_{\eta}^2v_\eta
+\frac{T_{\lambda_H}}{\sqrt{2}}v_{\bar{\eta}}v_S=0,
\\&&
\frac{1}{2}g_X^2(v_{\bar{\eta}}^2-v_{\eta}^2)v_{\bar{\eta}}+
\frac{1}{4}g_X(g_X+g_{YX})v_{\bar{\eta}}(v^2_{u}-v^2_{d})
+\frac{1}{2}\lambda_C^2v_{\bar{\eta}} (v_S^2+v_\eta^2)+l_W\lambda_Cv_{\eta}\nonumber\\&&
+\sqrt{2}M_S\lambda_Cv_Sv_{\eta}-\frac{1}{2}\lambda_H\lambda_Cv_{\eta}v_{u}v_d
+\frac{1}{2}\lambda_C\kappa v_{\eta} v_S^2+m_{\bar{\eta}}^2v_{\bar{\eta}}
+\frac{T_{\lambda_H}}{\sqrt{2}}v_{\eta}v_S=0,
\\&&(\lambda_H^2v_S+\sqrt{2}\mu\lambda_H)\frac{1}{2}v^2+\frac{1}{2}\lambda_C^2v_S\xi^2+4M_S^2v_S+\kappa^2v_S^3+ 2B_S v_S+\sqrt{2}L_{S}\nonumber\\&&+
2l_W(\sqrt{2}M_S+\kappa v_S)+\sqrt{2}M_S(\lambda_Cv_\eta v_{\bar\eta}-\lambda_H v_u v_d+3\kappa v_S^2)+m_S^2v_S
\nonumber\\&&+\lambda_C\kappa v_\eta v_{\bar{\eta}}v_S-\lambda_H\kappa v_u v_d v_S+\frac{1}{\sqrt{2}}(T_k v_S^2
+T_{\lambda_C} v_{\eta}v_{\bar{\eta}}-T_{\lambda_H} v_u v_d)=0.
\end{eqnarray}
The mass squared matrix for CP-odd Higgs in the basis $(\sigma_{d},\sigma_{u}, \sigma_{\eta}, \sigma_{\bar{\eta}}, \sigma_{s})$
 is diagonalized by $Z^A$. The neutral CP-even Higgs $\phi_d, \phi_u, \phi_\eta,\phi_{\bar{\eta}}$ and $\phi_S $ mix together at the tree level and they
form $5\times5$ mass squared matrix which is diagonalized by $Z^H$. Their concrete forms are collected in the Appendix.
As discussed in the MSSM, the loop corrections to the lightest CP-even Higgs mass are known to be large.
Therefore, we include the leading-log radiative corrections from stop and top particles \cite{LCTHiggs1,LCTHiggs2}.
The mass of the lightest Higgs boson can be written as
\begin{eqnarray}
&&m_h=\sqrt{(m_{h_1}^0)^2+\Delta m_h^2},\label{higgs mass}
\end{eqnarray}
with $m_{h_1}^0$ representing the lightest tree-level Higgs boson mass. The concrete form of $\Delta m_h^2$ is
\begin{eqnarray}
&&\Delta m_h^2=\frac{3m_t^4}{2\pi v^2}\Big[\Big(\tilde{t}+\frac{1}{2}+\tilde{X}_t\Big)+\frac{1}{16\pi^2}\Big(\frac{3m_t^2}{2v^2}-32\pi\alpha_3\Big)\Big(\tilde{t}^2
+\tilde{X}_t \tilde{t}\Big)\Big],\nonumber\\
&&\tilde{t}=\log\frac{M_{\tilde{T}}^2}{m_t^2},\qquad\;\tilde{X}_t=\frac{2\tilde{A}_t^2}{M_{\tilde{T}}^2}\Big(1-\frac{\tilde{A}_t^2}{12M_{\tilde{T}}^2}\Big).\label{higgs corrections}
\end{eqnarray}
$\alpha_3$ is the strong coupling constant. $M_{\tilde{T}}=\sqrt{m_{\tilde t_1}m_{\tilde t_2}}$
and $m_{\tilde t_{1,2}}$ are the stop masses. $\tilde{A}_t=A_t-\mu \cot\beta$ and $A_t$ is the trilinear Higgs stop coupling.

The neutrino mass matrix is deduced in the base $(\nu_L,\bar{\nu}_R)$
\begin{eqnarray}
M_{\nu}=
\left({\begin{array}{*{20}{c}}
0 & \frac{\upsilon_u}{\sqrt{2}}(Y_\nu^T)^{IJ}  \\
\frac{\upsilon_u}{\sqrt{2}}(Y_\nu)^{IJ} & \sqrt{2}\upsilon_{\bar{\eta}}(Y_X)^{IJ}  \\
\end{array}}
\right),
\end{eqnarray}
and it is diagonalized by the matrix $Z_\nu$ through the formula
\begin{eqnarray}
Z_\nu M_\nu Z^T_\nu=diag(M_\nu).
\end{eqnarray}

The mass matrix for CP-even sneutrino $({\phi}_{l}, {\phi}_{r})$ reads
\begin{eqnarray}
M^2_{\tilde{\nu}^R} = \left(
\begin{array}{cc}
m_{{\phi}_{l}{\phi}_{l}} &m^T_{{\phi}_{r}{\phi}_{l}}\\
m_{{\phi}_{l}{\phi}_{r}} &m_{{\phi}_{r}{\phi}_{r}}\end{array}
\right),\label{Rsneu}
 \end{eqnarray}
\begin{eqnarray}
&&m_{{\phi}_{l}{\phi}_{l}}= \frac{1}{8} \Big((g_{1}^{2} + g_{Y X}^{2} + g_{2}^{2}+ g_{Y X} g_{X})( v_{d}^{2}- v_{u}^{2})
+  g_{Y X} g_{X}(2 v_{\eta}^{2}-2 v_{\bar{\eta}}^{2})\Big)
\nonumber\\&&\hspace{1.8cm}+\frac{1}{2} v_{u}^{2}{Y_{\nu}^{T}  Y_\nu}  + m_{\tilde{L}}^2,
 \\&&m_{{\phi}_{l}{\phi}_{r}} = \frac{1}{\sqrt{2} } v_uT_\nu  +  v_u v_{\bar{\eta}} {Y_X  Y_\nu}
  - \frac{1}{2}v_d ({\lambda}_{H}v_S  + \sqrt{2} \mu )Y_\nu,\\&&
m_{{\phi}_{r}{\phi}_{r}}= \frac{1}{8} \Big((g_{Y X} g_{X}+g_{X}^{2})(v_{d}^{2}- v_{u}^{2})
+2g_{X}^{2}(v_{\eta}^{2}- v_{\bar{\eta}}^{2})\Big) + v_{\eta} v_S Y_X {\lambda}_{C}\nonumber \\&&\hspace{1.8cm}
 +m_{\tilde{\nu}}^2 + \frac{1}{2} v_{u}^{2}|Y_\nu|^2+  v_{\bar{\eta}} (2 v_{\bar{\eta}}Y_X  Y_X  + \sqrt{2} T_X).
\end{eqnarray}
To obtain the masses of sneutrinos, we use $Z^R$ to diagonalize $M^2_{\tilde{\nu}^R}$.

The mass matrix for CP-odd sneutrino $({\sigma}_{l}, {\sigma}_{r})$ is also deduced here
\begin{eqnarray}
M^2_{\tilde{\nu}^I} = \left(
\begin{array}{cc}
m_{{\sigma}_{l}{\sigma}_{l}} &m^T_{{\sigma}_{r}{\sigma}_{l}}\\
m_{{\sigma}_{l}{\sigma}_{r}} &m_{{\sigma}_{r}{\sigma}_{r}}\end{array}
\right),
 \end{eqnarray}
\begin{eqnarray}
&&m_{{\sigma}_{l}{\sigma}_{l}}= \frac{1}{8} \Big((g_{1}^{2} + g_{Y X}^{2} + g_{2}^{2}+  g_{Y X} g_{X})( v_{d}^{2}- v_{u}^{2})
+  2g_{Y X} g_{X}(v_{\eta}^{2}-v_{\bar{\eta}}^{2})\Big)
\nonumber\\&&\hspace{1.8cm}+\frac{1}{2} v_{u}^{2}{Y_{\nu}^{T}  Y_\nu}  + m_{\tilde{L}}^2,
 \\&&m_{{\sigma}_{l}{\sigma}_{r}} = \frac{1}{\sqrt{2} } v_uT_\nu -  v_u v_{\bar{\eta}} {Y_X  Y_\nu}
  - \frac{1}{2}v_d ({\lambda}_{H}v_S  + \sqrt{2} \mu )Y_\nu,\\&&
m_{{\sigma}_{r}{\sigma}_{r}}= \frac{1}{8} \Big((g_{Y X} g_{X}+g_{X}^{2})(v_{d}^{2}- v_{u}^{2})
+2g_{X}^{2}(v_{\eta}^{2}- v_{\bar{\eta}}^{2})\Big)- v_{\eta} v_S Y_X {\lambda}_{C}\nonumber \\&&\hspace{1.8cm}
+m_{\tilde{\nu}}^2 + \frac{1}{2} v_{u}^{2}|Y_\nu|^2+  v_{\bar{\eta}} (2 v_{\bar{\eta}}Y_X  Y_X  - \sqrt{2} T_X).
\end{eqnarray}
Using the matrix $Z^I$, we can diagonalize the mass squared matrix of the sneutrino $M^2_{\tilde{\nu}^I}$.
In the same way, we deduce the mass matrixes for slepton and neutralino, and show them in the Appendix.

Here, we show some needed couplings in this model.
The CP-odd Higgs bosons interact with $\tilde{\nu}^I$ and $\tilde{\nu}^R$, whose concrete form is
\begin{eqnarray}
&&\mathcal{L}_{A\tilde{\nu}^I\tilde{\nu}^R}=A_i\tilde{\nu}^I_j\frac{i}{4}\sum_{a,b=1}^3\Big\{\Big[2v_S\lambda_CZ^{R*}_{k3+b}Z^{I*}_{j3+a}(Y_X)_{ab}Z^A_{i3}-2\sqrt{2}Z^{R*}_{kb}Z^{I*}_{j3+a}(T_\nu)_{ab}Z^A_{i2}\nonumber\\&&
-2\sqrt{2}Z^{R*}_{k3+b}Z^{I*}_{j3+a}(T_X)_{ab}Z^A_{i4}+2v_\eta\lambda_CZ^{R*}_{k3+b}Z^{I*}_{j3+a}(Y_X)_{ab}Z^A_{i5}\Big]+\Big[R\leftrightarrow I, j\leftrightarrow k\Big]
\Big\}\tilde{\nu}^{*R}_k.
\end{eqnarray}
We also deduce the vertexes of $\tilde{\nu}^R_k-\bar{e}_i-\chi_j^-$ and $\tilde{\nu}^R_k-\nu_i-\bar{\chi}_i^0$,
\begin{eqnarray}
&&\mathcal{L}_{\tilde{\nu}^R\bar{e}\chi^-}=\bar{e}_i\Big\{\frac{i}{\sqrt{2}}U^*_{j2}Z^{R*}_{ki}Y_e^iP_L-\frac{i}{\sqrt{2}}g_2V_{j1}Z^{R*}_{ki}P_R\Big\}\chi_j^-\tilde{\nu}^R_k,
\\&&\mathcal{L}_{\tilde{\nu}^R\nu\bar{\chi}^0}=\bar{\chi}_i^0\Big\{\frac{i}{2}(-g_2Z^{N*}_{i2}+g_{YX}Z^{N*}_{i5}+g_1Z^{N*}_{i1})
\sum_{a=1}^3Z^{R*}_{ka}U_{ja}^{V*}P_L\nonumber\\&&\hspace{1.7cm}+
\frac{i}{2}(-g_2Z^N_{i2}+g_{YX}Z^N_{i5}+g_1Z^N_{i1})\sum_{a=1}^3Z^{R*}_{ka}U_{ja}^{V}P_R\Big\}\nu_i\tilde{\nu}^R_k.
\end{eqnarray}
To save space in the text, the remaining vertexes are placed in Appendix.

\section {relic density}

In this section, we suppose the lightest mass eigenstate $(\tilde{\nu}^R_1)$ of CP-even sneutrino mass squared matrix in Eq. (\ref{Rsneu}) as a dark matter candidate and
calculate the relic density. Any WIMP candidate has to satisfy the relic density constraints.
The $\tilde{\nu}^R_1$ number density $n_{\tilde{\nu}^R_1}$ is governed by the Boltzmann  equation \cite{rotation1,boltzmann11,boltzmann12,XFBO1}
\begin{eqnarray}
\frac{d n_{\tilde{\nu}_1^R}}{dt}=-3Hn_{\tilde{\nu}^R_1}-\langle\sigma v\rangle_{SA}(n^2_{\tilde{\nu}^R_1}-n^2_{\tilde{\nu}^R_1 eq})
-\langle\sigma v\rangle_{CA}(n_{\tilde{\nu}^R_1}n_\phi-n_{\tilde{\nu}^R_1 eq}n_{\phi eq}).
\end{eqnarray}
 $\tilde{\nu}^R_1$ can both self-annihilate and co-annihilate with another specy $\phi$. When the
 annihilation rate of  $\tilde{\nu}^R_1$ becomes roughly equal to the Hubble expansion rate,
 the species freeze out at the temperature $T_F$,
\begin{eqnarray}
\langle\sigma v\rangle_{SA}n_{\tilde{\nu}^R_1}+\langle\sigma v\rangle_{CA}n_{\phi}\sim H(T_F).
\end{eqnarray}
With the supposition $M_\phi>M_{\tilde{\nu}^R_1}$ \cite{importantGS}
\begin{eqnarray}
n_\phi=\Big(\frac{M_\phi}{M_{\tilde{\nu}^R_1}}\Big)^{3/2}\texttt{Exp}[(M_{\tilde{\nu}^R_1}-M_\phi)/T]n_{\tilde{\nu}^R_1}.
\end{eqnarray}
Then it becomes
\begin{eqnarray}
\Big[\langle\sigma v\rangle_{SA}+\langle\sigma v\rangle_{CA}
\Big(\frac{M_\phi}{M_{\tilde{\nu}^R_1}}\Big)^{3/2}\texttt{Exp}[(M_{\tilde{\nu}^R_1}-M_\phi)/T]\Big]n_{\tilde{\nu}^R_1}\sim H(T_F).
\end{eqnarray}

We study its annihilation rate $\langle\sigma v\rangle_{SA}$ ($\langle\sigma v\rangle_{CA}$)
and its relic density $\Omega_D$ in the thermal history of the universe. To this end, the self-annihilation cross section $\sigma(\tilde{\nu}^R_1 \tilde{\nu}_1^{R*} \rightarrow$ anything) and
co-annihilation cross section $\sigma(\tilde{\nu}^R_1 \phi \rightarrow$ anything) should be calculated.  In the center of mass frame,  their results can be written as
$\sigma v_{rel}=a+bv_{rel}^2$, with $v_{rel}$ denoting the relative velocity of the two particles in the initial states.
It is a good approximation to calculate the freeze-out temperature ($T_F$) from the following formula \cite{rotation1,HXG,XFCW,XFBO1}
\begin{eqnarray}
&&x_F=\frac{m_D}{T_F}\simeq\ln[\frac{0.038M_{Pl}m_D(a+6b/x_F)}{\sqrt{g_*x_F}}].
\end{eqnarray}
$M_{Pl}$ is the Planck mass $1.22\times10^{19}$ GeV.  $m_D=m_{\tilde{\nu}^R_1}$ denoting the WIMP mass and $x_F\equiv m_D/T_F$.
 $g_*$ is the number of the relativistic degrees of freedom with mass less than $T_F$.
 The formula for the density of cold non-baryonic matter can be simplified in the following form \cite{rotation1,longlife1,XFBO1,zhaosm}
\begin{eqnarray}
\Omega_D h^2\simeq \frac{1.07\times10^9 x_F}{\sqrt{g_*}M_{PL}(a+3b/x_F)~\rm{GeV} }\;,
\end{eqnarray}
and its value should be $\Omega_D h^2=0.1186\pm 0.0020$ \cite{pdg}.

The dominant processes for the self-annihilation are: $\tilde{\nu}^R_1+\tilde{\nu}^R_1\rightarrow \{(W+W),~(Z+Z),~(h+h),~(\bar{u}_i+ u_i),
~(\bar{d}_i+ d_i),~(\bar{l}_i+ l_i),~(\bar{\nu}_i+ \nu_i)\}$ with $i=1,~2,~3$, $h$ representing the lightest CP-even Higgs. $\nu_i$ denote
three light neutrinos.
The studied co-annihilation processes read as:

a. $\tilde{\nu}^R_1+\tilde{\nu}^R_k\rightarrow \{(W+W),~(Z+Z),~(h+h),~(\bar{u}_i+ u_i),
~(\bar{d}_i+ d_i),~(\bar{l}_i+ l_i),~(\bar{\nu}_i+ \nu_i)\}$ with $k=2\dots6,~i=1,~2,~3$.

b. $\tilde{\nu}^R_1+\tilde{\nu}^I_j\rightarrow \{(W+W),~(Z+h),~(\bar{u}_i+ u_i),
~(\bar{d}_i+ d_i),~(\bar{l}_i+ l_i),~(\bar{\nu}_i+ \nu_i)\} $ and $j=1\dots6,~i=1\dots3$.

c. $\tilde{\nu}^R_1+\chi^0_n\rightarrow \{(Z+\nu_i),~(W^++l_i^-), ~(W^-+l_i^+)\}$ and $n=1\dots8,~i=1\dots3$.

\section{direct detection}

   The main scattering processes of CP-even sneutrinos off nucleons are $\tilde{\nu}^R+q\rightarrow \tilde{\nu}^R+ q$
   and $\tilde{\nu}^R +q\rightarrow \tilde{\nu}^I +q$.  For the first type process $\tilde{\nu}^R+q\rightarrow \tilde{\nu}^R+ q$,
   the exchanged particles are CP-even Higgs. While, for the second type process $\tilde{\nu}^R +q\rightarrow \tilde{\nu}^I +q$,
    the exchanged particles are vector bosons $Z$ and $Z^\prime$.
The CP-odd Higgs boson contributions are much smaller than the contributions from CP-even Higgs boson and can be neglected safely \cite{LJandHE}.
After some calculation, we obtain the operators $\tilde{\nu}^{R*}\tilde{\nu}^R \bar{q}q$ and $\tilde{\nu}^{R*} \partial_\mu \tilde{\nu}^R \bar{q}\gamma^\mu q$ at the quark level.

To get the final results, we should convert the quark level coupling to the effective nucleon coupling.
For the operator $\tilde{\nu}^{R*}\tilde{\nu}^R \bar{q}q$, the useful expressions are shown below \cite{LJandHE}
\begin{eqnarray}
&&a_qm_q\bar{q}q\rightarrow f_Nm_N\bar{N}N,~~~~~~~~~~
f_N=\sum_{q=u,d,s}f_{Tq}^{(N)}a_q+\frac{2}{27}f_{TG}^{(N)}\sum_{q=c,b,t}a_q,\nonumber\\&&\langle N|m_q \bar{q}q|N\rangle=m_N f_{Tq}^{(N)},
~~~~~~~f_{TG}^{(N)}=1-\sum_{q=u,d,s}f_{Tq}^{(N)}.
\end{eqnarray}
$f_N$ includes the coupling to gluons induced by integrating out heavy quark loops.
The numbers of $f_{Tq}^{(N)}$ are collected here \cite{DarkSUSY1,DarkSUSY2,DarkSUSY3},
\begin{eqnarray}
&&f^{(p)}_{Tu}=0.0153,~~~f^{(p)}_{Td}=0.0191,~~~f^{(p)}_{Ts}=0.0447, \nonumber\\&&
f^{(n)}_{Tu}=0.0110,~~~f^{(n)}_{Td}=0.0273,~~~f^{(n)}_{Ts}=0.0447.
\end{eqnarray}

It is easy to convert the operator $b_q\tilde{\nu}^{R*} \partial_\mu \tilde{\nu}^R \bar{q}\gamma^\mu q$
to $b_N \tilde{\nu}^{R*} \partial_\mu \tilde{\nu}^R \bar{N}\gamma^\mu N$ through the following formulas
\begin{eqnarray}
 b_p=2b_u+b_d,~~~b_n=2b_d+b_u.
\end{eqnarray}

With the obtained $f_N$, one gets the scattering cross section
\begin{eqnarray}
\sigma=\frac{1}{\pi}\mu^2[Z_pf_p+(A-Z_p)f_n]^2.
\end{eqnarray}
Here $Z_p$ is the number of proton, and $A$ represents the number of atom.
\section{numerical results}

In this section, we study the numerical results.
$Z'$ boson properties are constrained by manifold low energy experiments \cite{low energy1, low energy2}. The lower limits on the mass of
$Z'$ set by low energy data are about 1 TeV in some models.  The mass bounds for $M_{Z'}$ from LHC are about several TeV, which are more
severe than those from low energy constraints.
In the case of final
states with taus, the lower mass limits for $Z^\prime$ obtained at 13 TeV are as high as $\sim$ 2.4 TeV \cite{ZP1}.
Another stringent for the mass of $Z^\prime$  is set in the fully hadronic channel, with a lower
mass limit of 2.35 TeV in the context of the Heavy Vector Triplet model weakly -coupled scenario A \cite{ZP2}.
The result from ATLAS Collaboration at $\sqrt{s}=13$ TeV obtained with 2016 data is more stringent \cite{ATLAS2016}.
The resulting $95\% $ CL lower mass limits are 4.5 TeV for the $Z^\prime _{SSM}$ in the Sequential Standard Model, 4.1 TeV for the
 $Z^\prime_{\chi}$, and 3.8 TeV for the $Z^\prime_{\psi}$. Here, $Z^\prime_{\chi}$ and $Z^\prime_{\psi}$ belong to the $E_6$-motivated model.
 Other $E_6$ $Z^\prime$ models are also constrained in the range between those quoted for the $Z^\prime_\chi$ and $Z^\prime_{\psi}$.
 The lower mass limits are 4.1 TeV for the $Z^\prime_{3R}$ in the left-right symmetric model, and 4.2 TeV for the $Z^\prime_{B-L}$ of the (B-L) model \cite{ATLAS2016}.
 The authors \cite{ZPG1,ZPG2} give the upper bound($M_{Z^\prime}/g_X\geq6$ TeV) on the ratio between $M_{Z^\prime}$ and its gauge
coupling at 99\% CL. $\tan \beta_\eta$ is also constrained by the LHC experimental data and should be smaller than 1.5 \cite{TanBP}.
In order to satisfy the constraints from LHC, we choose the parameters to make $M_{Z^{\prime}}> 4.5$ TeV,
because the quoted number are valid in other models and do not apply directly.
The constraints for supersymmetric particles, shown in Ref. \cite{pdg}, are also taken into account.

Considering the above constraints, we use the following parameters
\begin{eqnarray}
&&M_S =0.8 ~{\rm TeV},~
T_{\kappa} =1.6~ {\rm TeV}, ~
M_1 =M_2 = M_{BL}=1~{\rm TeV},
~\tan\beta = 11,~g_{YX}=0.2,\nonumber\\&&
\upsilon_{\eta} = 15.5\times\cos\beta_\eta ~{\rm TeV},~
\upsilon_{\bar{\eta}} = 15.5\times\sin\beta_\eta~{\rm TeV},~Y_{X11} =Y_{X22} = 0.5,~Y_{X33} =0.4,
\nonumber\\&&
g_X=\kappa=\lambda_H = 0.3,~\lambda_C = -0.3,~ M^2_Q=2.5~{\rm TeV}^2,~ M_{BB^\prime}=0.4~{\rm TeV},
~T_{\lambda_H} = 1.8~{\rm TeV},\nonumber\\&&
T_{X11} =T_{X22} = -1~{\rm TeV},~T_{X33} = -2~{\rm TeV}, ~
T_{e11} =T_{e22} = -3~{\rm TeV},~T_{e33} = -4~{\rm TeV},
\nonumber\\&& l_W = 4~{\rm TeV}^2,
~ B_{\mu} = B_S=M^2_T=1~{\rm TeV}^2,~\tan\beta_\eta=0.83,~T_{\nu11} = T_{\nu22}=0,~ A_t=2.6{\rm TeV},
\nonumber\\&&
T_{\lambda_C} = 0.25~{\rm TeV},~
M^2_{\nu11} = M^2_{\nu22}=0.5~{\rm TeV}^2,~ M^2_{L11}= M^2_{L22} = M^2_{E11}= M^2_{E22} = 3~{\rm TeV}^2.
\end{eqnarray}
Here, we take $T_\nu,~T_X $ and $M_\nu$ as diagonal matrices, for example
\begin{eqnarray}
T_X = \left(
\begin{array}{ccc}
T_{X11} & 0 & 0\\
0 &T_{X22} & 0\\
0 & 0 &T_{X33}
\end{array}
\right).
\end{eqnarray}
We list the remaining parameters which will vary in the following numerical analysis:
\begin{eqnarray}
v_S,~~ \mu, ~~T_{\nu33},~~ M^2_{L33}, ~~M^2_{\nu33},~~ M^2_{E33}, ~~m_S^2.
\end{eqnarray}

Firstly, we research the lightest CP-even Higgs mass including the loop corrections and discuss the other CP-even  Higgs masses.
Secondly, the relic density of the lightest CP-even sneutrino is calculated numerically.
At last, we study the cross section for the lightest sneutrino scattering off nucleon.

\subsection{ Higgs mass}
\begin{figure}[h]
\setlength{\unitlength}{1mm}
\centering
\includegraphics[width=3.6in]{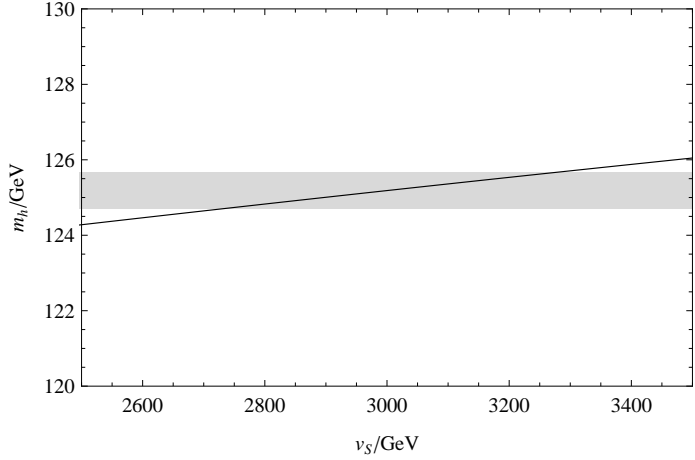}
\caption[]{ Considering the loop corrections, the lightest
CP-even Higgs mass ($m_h$) versus $v_{S}$ is plotted by the solid line with $A_t$=2.6 TeV.\label{vs}}
\end{figure}

Considering the loop corrections from top and stop contributions, we study the SM-like Higgs boson mass in this subsection.
For simplicity, we suppose that $\mu=0.5 ~{\rm TeV}$ and $m_S^2=1 ~{\rm TeV}^2$ in the following analysis.
$v_S$ is the VEV of $S$ and  emerges in the diagonal elements of CP-even Higgs mass squared matrix (Eqs. (A1) and (A2)).
So, the lightest tree-level Higgs mass $m_{h_1}^{0}$ is the increasing function of $v_S$. More important, $v_S$ affects the lightest
neutralino mass through the element $m_{\tilde{H}_d^0\tilde{H}_u^0} = - \frac{1}{\sqrt{2}} {\lambda}_{H} v_S  - \mu$ in Eq. (A7). In our
used parameter space, to keep $\tilde{\nu}_1^R$ as LSP we run $v_S$ from 2500 to 3500 GeV and
 show $m_h$ varying with $v_S$ in  Fig. \ref{vs}.  The gray area is the lightest CP-even Higgs mass in $\pm 3\sigma$
  sensitivity band. Obviously, $m_h$ is the increasing
function of $v_S$. In the $v_S$ region (2700-3300) GeV, $m_h$ can satisfy the experimental bound on the SM-like Higgs boson mass in $\pm3\sigma$ sensitivity.
And the other CP-even Higgs boson masses are all heavier than 2.5 TeV in this case.

\subsection{Relic density of sneutrino dark matter}
\begin{figure}[h]
\setlength{\unitlength}{1mm}
\centering
\includegraphics[width=2.9in]{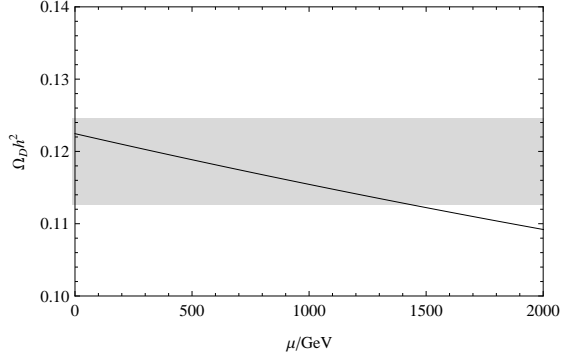}
\caption[]{The relic density  versus $\mu$.}\label{fdmutu}
\end{figure}

Here, the parameters $v_S=3~{\rm TeV}$ and $m_S^2=1~ {\rm TeV}^2$ are used to study the
relic density of dark matter.
With the same parameters, the lightest neutralino in the MSSM is around 500 GeV as $\mu=500$ GeV. When $|\mu|$ is near zero, the mass of the lightest
neutralino in the MSSM is very tiny. However, the case in the $U(1)_X$SSM is different from that in the MSSM.
The neutralino mass matrix (Eqs. (A7)) in the $U(1)_X$SSM is $8\times8$, where
$m_{\tilde{H}_d^0\tilde{H}_u^0} = - \frac{1}{\sqrt{2}} {\lambda}_{H} v_S  - \mu$ (Eqs. (A8)) corresponds to $-\mu$ in the
neutralino mass matrix of MSSM. According to our parameters $v_S=3000 ~{\rm GeV}$ and $\lambda_H=0.3$,
$m_{\tilde{H}_d^0\tilde{H}_u^0} = - \frac{1000}{\sqrt{2}}{\rm GeV}- \mu\sim-707{\rm GeV}-\mu$. That is to say, $m_{\tilde{H}_d^0\tilde{H}_u^0}$
is equal to the shift of $-\mu$. Therefore, as $\mu=0$, $m_{\tilde{H}_d^0\tilde{H}_u^0}$ is around -707 GeV, and the lightest neutralino
is around 650 GeV. The other terms in Eqs. (A7) slightly influence the lightest neutralino.
When $\mu$ is during the region $(-400, -1100)$ GeV, the lightest neutralino will be small,
but in this parameter space the corresponding relic density $\Omega_D h^2$ can not satisfy the non-baryonic density value.
Considering these constraints, we plot $\Omega_D h^2$ versus $\mu$ in Fig. \ref{fdmutu} with $\mu$ varying from 0 to 2000 GeV. The remaining parameters are
 $m^2_{\nu33}=250^2~{\rm GeV}^2,~T_{\nu33}=1.6~{\rm TeV},~ M^2_{L33}=M^2_{E33}=3~{\rm TeV}^2.$
 The gray area represents the relic density in $\pm 3\sigma$
  sensitivity band.
 In the $\mu$ region (0, 2000) GeV, the relic density is the decreasing function. From this diagram, one can find that as $\mu$ near
 500 GeV the result is close to the center value of the relic density.
In $\pm 1\sigma$ sensitivity of $\Omega_D h^2$, the lightest neutralino is around 850 GeV. The choice of parameters are chosen
for illustration.

\begin{figure}[h]
\setlength{\unitlength}{1mm}
\centering
\includegraphics[width=2.9in]{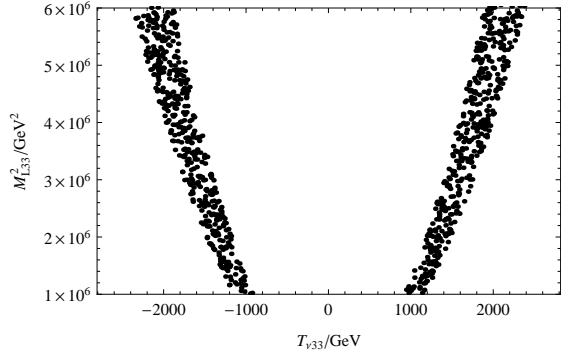}
\caption[]{The allowed results of the relic density in the plane of $M^2_{L33}$ and $T_{\nu33}$}\label{tuMLTN}
\end{figure}

To more accurately scan the parameter space,
the numerical results of the relic density in $\pm 3\sigma$ sensitivity are plotted
in the plane of $M^2_{L33}$ and $T_{\nu33}$ as $\mu=500~{\rm GeV},~m^2_{\nu33}=250^2~{\rm GeV}^2,~M^2_{E33}=3~{\rm TeV}^2$. $M_{L33}^2$ and $T_{\nu33}$ come from the soft breaking terms.
As the non-diagonal element of sneutrino mass matrix, $T_{\nu33}$ affects the sneutrino masses and mixing.
On the other hand, $M_{L33}^2$ appears in the diagonal elements of the mass matrixes for sneutrino and slepton. So,
$M_{L33}^2$ influences the both type scalars.
 The allowed results are plotted by the dots in  Fig. \ref{tuMLTN}, where
they are almost symmetric with respect to $T_{\nu33}=0$.

\begin{figure}[h]
\setlength{\unitlength}{1mm}
\centering
\includegraphics[width=2.9in]{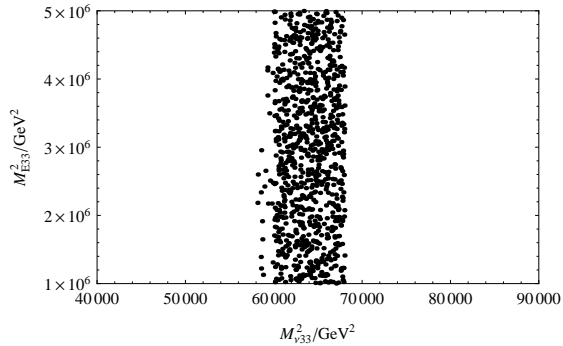}
\caption[]{The allowed results of the relic density in the plane of $M^2_{E33}$ and $M^2_{\nu33}$}\label{tuMEMN}
\end{figure}

In the plan of $M^2_{E33}$ and $ M^2_{\nu33}$,
the allowed results  in $\pm 3\sigma$
  sensitivity of $\Omega_Dh^2$
 are also researched
by taking $\mu=500~{\rm GeV},~T_{\nu33}=1.6~{\rm TeV}$ and $ M^2_{L33}=3~{\rm TeV}^2$.
We show these results by the dots in Fig. \ref{tuMEMN}.
The effect of $M^2_{E33}$ is small, because it influences the numerical results only by affecting the slepton mixing and masses.
$M^2_{\nu33}$ appears in the mass matrix of sneutrino, which can affect the lightest sneutrino mass and the mixing of sneutrino.
  Therefore, $M_{\nu33}^2$ is a sensitive parameter, and has obvious influence on $m_{\tilde{\nu}_1^R}$ and $\Omega_Dh^2$.
 The favorite region of $M^2_{\nu33}$ is from 60000 to 68000 ${\rm GeV}^2$. This region of $M^2_{\nu33}$ can also keep
the lightest CP-even sneutrino $\tilde{\nu}_1^R$ as LSP.

 According to the parameter space under consideration, the lightest CP-even sneutrino mass is about 320 GeV.
   The other CP-even sneutrinos ($\tilde{\nu}^R_2\dots\tilde{\nu}^R_6$) are all heavier
than 1900 GeV. The masses of all CP-odd sneutrinos ($\tilde{\nu}^I_1\dots\tilde{\nu}^I_6$) are larger than 1900 GeV.
For the relic density in  $\pm 1\sigma$ sensitivity, the lightest neutralino is around 850 GeV.
That is to say  $\tilde{\nu}^R_1$ is the LSP, and can be the dark matter candidate.

If the mass of the virtual particle in  s-channel is around $2M_D$, the resonance
annihilation will occur. The resonance annihilation strongly
affects the annihilation cross-section hence the relic density.
In these numerical results, the mass of dark matter
is $M_D\sim 320$ GeV. The four virtual CP-even Higgs bosons in s-channel are all heavier than 2.5 TeV, and the lightest CP-even
Higgs boson is about 125 GeV. It is obvious that $2M_D$ is far from all the CP-even Higgs masses. So the resonance annihilation
can not take place.

\subsection{The cross section of the sneutrino scattering off nucleon }

Taking into account the constraint from the relic density, we calculate numerically the cross section of the sneutrino scattering off nucleon in this subsection.
 Within the considered parameter space, the lightest CP-even sneutrino
is around 320 GeV.
The experimental limit on direct detection for
a dark matter of 320 GeV is about $2.5\times10^{-46}~{\rm cm}^2$ for Xenon and about twice as large
for PandaX \cite{PanXen1,PanXen2}.
Using the parameters $v_S=3 ~{\rm TeV},~ m^2_{\nu33}=250^2~{\rm GeV}^2,~ M^2_{L33}=M^2_{E33}=3~{\rm TeV}^2$ that can satisfy
the relic density constraint, we research the cross section of the sneutrino scattering off nucleon.

\begin{figure}[h]
\setlength{\unitlength}{1mm}
\centering
\includegraphics[width=3.6in]{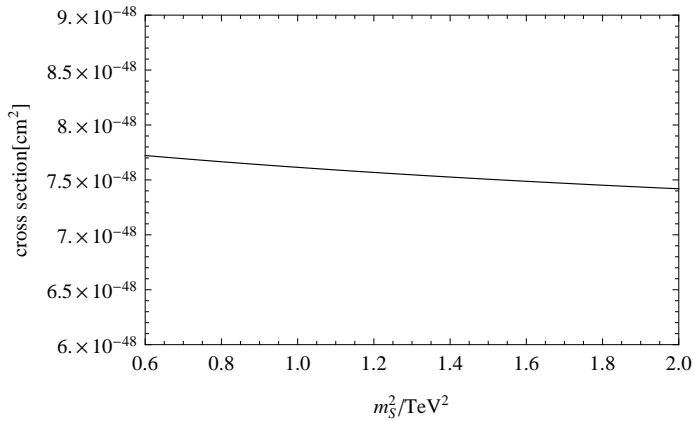}
\caption[]{The cross section versus $m_S^2$. }\label{tuhemsf}
\end{figure}

$m_S^2$ is the mass square term of $S^2$ in the soft breaking terms. It does not have relation with the masses of sneutrinos and neutralinos.
Because of the mixing of S and neutral CP-even Higgs (Eqs. (A1) and (A2)), $m_S^2$ impacts the CP-even Higgs masses and Higgs mixing to some extent.
 $m_S^2$ can directly improve heavy Higgs mass, but its effect to the lightest CP-even Higgs mass $m^0_{h_1}$ at the tree-level is very small.
CP-even Higgs bosons give dominant contribution to the relic density, so $m_S^2$ is constrained by $\Omega_Dh^2$. Considering this constraint,
we adopt  $m_S^2$ region as $[0.6, 2.0] ~ {\rm TeV}^2$.
In Fig. \ref{tuhemsf}, the cross section versus $m_S^2$ is plotted by the solid line with $\mu=500~{\rm GeV}$ and $T_{\nu33}=1.6~{\rm TeV}$. The solid line is in the region
$(6.5\times 10^{-48},~8.0\times 10^{-48})~{\rm cm}^2$,  when $m_S^2$ varies from $0.6$ to $2~ {\rm TeV}^2$.
These results for $m_D\sim 320$ GeV are more than one order of magnitude below current limits.

\begin{figure}[h]
\setlength{\unitlength}{1mm}
\centering
\includegraphics[width=3.6in]{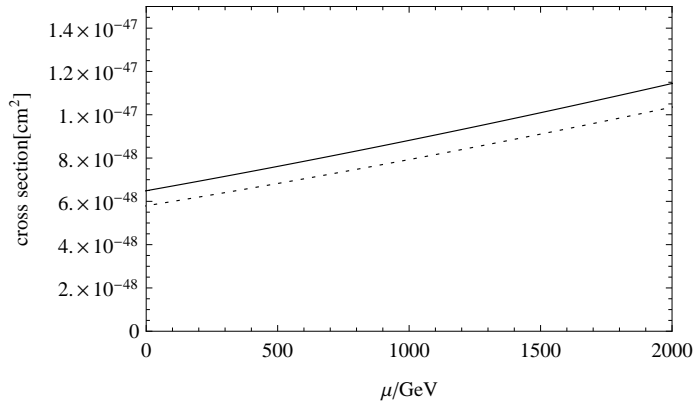}
\caption[]{The cross section versus $\mu$ is plotted by
solid line (dotted line) with $T_{\nu33}=1.6~ (1.4)$ TeV. }\label{tuhemu}
\end{figure}

To further discuss the sneutrino scattering off nucleon, in Fig. \ref{tuhemu} we plot the cross section versus $\mu$ by the
solid line (dotted line) with $T_{\nu33}=1.6~ (1.4)$ TeV and $m_S^2=3 ~{\rm TeV^2}$.
 As discussed for the Fig. \ref{fdmutu}, to satisfy the constraints from $\Omega_D h^2$ and $\tilde{\nu}_{1}^R$ as the LSP, we take $\mu$ in the region [0, 2000] GeV.
 For the same $\mu$, the value of the solid line is a little bigger than the value of the dotted line.
The solid line and dotted line possess similar behaviors and they are increasing functions of $\mu$.
As $\mu=500$ GeV, the solid line and dotted line are around $7\times10^{-48}~{\rm cm^2}$. While, the cross section can reach $10^{-47}~{\rm cm^2}$ with $\mu$ near 2000 GeV.
In our parameter space, the theoretical predictions for this model for the
benchmark chosen are smaller than the current limits by one order of magnitude.

\section{discussion and conclusion}
The $U(1)_X$SSM is the extension of MSSM, whose local
gauge group is $SU(3)_C\times SU(2)_L \times U(1)_Y\times U(1)_X$.
To obtain this model, righ-handed neutrinos and three Higgs superfields $\hat{\eta},~\hat{\bar{\eta}},~\hat{S}$ are added to the MSSM.
Through the seesaw mechanism, three tiny neutrino masses can be produced. The right-handed sneutrinos are sterile, and if they
are main parts of the lightest sneutrino, it possesses the characters of cold dark matter.

Taking into account the loop corrections, we study the lightest CP-even Higgs mass (SM-like) in the $U(1)_X$SSM. Comparing with the MSSM, there are three additional Higgs superfields($\hat{\eta},~\hat{\bar{\eta}},~\hat{S}$) in the $U(1)_X$SSM, which is also discussed.
 With the assumption that the lightest CP-even sneutrino can be a cold dark matter
candidate, the relic density of dark matter and the cross section of dark matter scattering off nucleon are both studied.
The virtual Higgs contributions to both the relic density and the scattering cross section are dominant.
 The numerical results imply that the parameters $M^2_{\nu33},M^2_{L33},T_{\nu33}$ and $\mu$ are all important.
 The used parameter space is reasonable and satisfy the dark matter constraints from both the relic density
and the scattering off nucleon.  This work gives constraints to the parameter space of the $U(1)_X$SSM and may be benefit for the future direct detection.

{\bf Acknowledgments}

We are very grateful to Wei Chao the professor of Beijing Normal University for giving us some useful discussions and Tiago Adorno the professor of Hebei University for English rewriting.
This work is supported by National Natural Science Foundation of China (NNSFC) (No. 11535002, No. 11605037, No. 11705045),
Post-graduate's Innovation Fund Project of Hebei Province
 (No. CXZZBS2019027),  Hebei Key Lab of Optic-Electronic Information and Materials, and the youth top-notch talent support program of the Hebei Province.

\appendix
\section{mass matrix}

In the basis $(\phi_{d}, \phi_{u}, \phi_{\eta}, \phi_{\bar{\eta}}, {\phi}_{s})$, the mass squared matrix of CP-even Higgs reads
\begin{eqnarray}
m^2_{h} = \left(
\begin{array}{ccccc}
m_{\phi_{d}\phi_{d}} &m_{\phi_{u}\phi_{d}} &m_{\phi_{\eta}\phi_{d}} &m_{\phi_{\bar{\eta}}\phi_{d}} &m_{{\phi}_{s}\phi_{d}}\\
m_{\phi_{d}\phi_{u}} &m_{\phi_{u}\phi_{u}} &m_{\phi_{\eta}\phi_{u}} &m_{\phi_{\bar{\eta}}\phi_{u}} &m_{{\phi}_{s}\phi_{u}}\\
m_{\phi_{d}\phi_{\eta}} &m_{\phi_{u}\phi_{\eta}} &m_{\phi_{\eta}\phi_{\eta}} &m_{\phi_{\bar{\eta}}\phi_{\eta}} &m_{{\phi}_{s}\phi_{\eta}}\\
m_{\phi_{d}\phi_{\bar{\eta}}} &m_{\phi_{u}\phi_{\bar{\eta}}} &m_{\phi_{\eta}\phi_{\bar{\eta}}} &m_{\phi_{\bar{\eta}}\phi_{\bar{\eta}}} &m_{{\phi}_{s}\phi_{\bar{\eta}}}\\
m_{\phi_{d}{\phi}_{s}} &m_{\phi_{u}{\phi}_{s}} &m_{\phi_{\eta}{\phi}_{s}} &m_{\phi_{\bar{\eta}}{\phi}_{s}} &m_{{\phi}_{s}{\phi}_{s}}\end{array}
\right).\label{CPevenH}
 \end{eqnarray}
The explicit forms of the elements $m_{\phi_{d}\phi_{d}}$ etc in this mass matrix are shown
{\small\begin{eqnarray}
&&m_{\phi_{d}\phi_{d}}= m_{H_d}^2+  |\mu|^2
 +\frac{1}{8} \Big( [g_{1}^{2}+(g_{X}+g_{YX})^{2}+g_2^2] (3 v_{d}^{2}  - v_{u}^{2})
 \nonumber \\ &&+2 (g_{Y X} g_{X}+g_X^2) ( v_{\eta}^{2}- v_{\bar{\eta}}^{2})\Big)
 +  \sqrt{2} v_S \mu {\lambda}_{H}   +\frac{1}{2} (v_{u}^{2} + v_S^{2})|{\lambda}_{H}|^2,\nonumber
 \\&&m_{\phi_{d}\phi_{u}} = -\frac{1}{4} \Big(g_{2}^{2} + (g_{Y X} + g_{X})^{2} + g_1^{2}\Big)v_d v_u
 + |{\lambda}_{H}|^2 v_d v_u - {\lambda}_{H} l_W \nonumber \\&&-\frac{1}{2}{\lambda}_{H} (v_{\eta} v_{\bar{\eta}} {\lambda}_{C}  + v_S^{2} \kappa )
 - B_{\mu}- \sqrt{2} v_S (\frac{1}{2}T_{{\lambda}_{H}}  + M_S {\lambda}_{H} ),\nonumber
 \\ &&m_{\phi_{u}\phi_{u}} = m_{H_u}^2+ |\mu|^2+\frac{1}{8} \Big( [g_{1}^{2}+(g_{X}+g_{YX})^{2}+g_2^2] (3 v_{u}^{2}  - v_{d}^{2})
 \nonumber \\ &&+2 (g_{Y X} g_{X}+g_X^2) ( v_{\bar{\eta}}^{2}-v_{\eta}^{2})\Big)
 +  \sqrt{2} v_S\mu {\lambda}_{H}   + \frac{1}{2}(v_{d}^{2} + v_S^{2})|{\lambda}_{H}|^2,\nonumber
 \\&&m_{\phi_{d}\phi_{\eta}} = \frac{1}{2}g_{X} (g_{Y X} + g_{X})v_d v_{\eta}
  -\frac{1}{2} v_u v_{\bar{\eta}} {\lambda}_{H} {\lambda}_{C} ,\nonumber\\&&
m_{\phi_{u}\phi_{\eta}} = -\frac{1}{2}g_{X} (g_{Y X} + g_{X})v_u v_{\eta}
-\frac{1}{2} v_d v_{\bar{\eta}} {\lambda}_{H} {\lambda}_{C}, \nonumber\\&&
m_{\phi_{\eta}\phi_{\eta}} = m_{\eta}^2 +\frac{1}{4} \Big((g_{Y X} g_{X}+g_X^2) ( v_{d}^{2}
- v_{u}^{2})+2g_{X}^{2}
( 3 v_{\eta}^{2}-v_{\bar{\eta}}^{2})\Big)+\frac{|{\lambda}_{C}|^2}{2} (v_{\bar{\eta}}^{2} + v_S^{2}), \nonumber\\&&
m_{\phi_{d}\phi_{\bar{\eta}}} = -\frac{1}{2}g_{X} (g_{Y X} + g_{X})v_d v_{\bar{\eta}}
  -\frac{1}{2} v_u v_{\eta} {\lambda}_{H} {\lambda}_{C},\nonumber \\&&
m_{\phi_{u}\phi_{\bar{\eta}}} = \frac{1}{2}g_{X} (g_{Y X} + g_{X})v_u v_{\bar{\eta}}  -\frac{1}{2} v_d v_{\eta} {\lambda}_{H} {\lambda}_{C},\nonumber\\&&
m_{\phi_{\eta}\phi_{\bar{\eta}}} = - g_{X}^{2}v_{\eta} v_{\bar{\eta}}+\frac{1}{2}(2 l_W  - {\lambda}_{H} v_d v_u )
{\lambda}_{C} + |{\lambda}_{C}|^2 v_{\eta} v_{\bar{\eta}}\nonumber \\
 && + \frac{1}{\sqrt{2}} v_S (2 M_S {\lambda}_{C} + T_{{\lambda}_{C}}) + \frac{1}{2} v_S^{2} {\lambda}_{C} \kappa,\nonumber
\\&&
m_{\phi_{\bar{\eta}}\phi_{\bar{\eta}}} = m_{\bar{\eta}}^2+\frac{1}{4} \Big((g_{Y X} g_{X}+g_X^2)
 ( v_{u}^{2}- v_{d}^{2})+2g_{X}^{2}( 3 v_{\bar{\eta}}^{2}-v_{\eta}^{2})\Big)+\frac{|{\lambda}_{C}|^2 }{2} \Big(v_{\eta}^{2} + v_S^{2}\Big),\nonumber \\&&
m_{\phi_{d}{\phi}_{s}} = \Big({\lambda}_{H} v_d v_S  + \sqrt{2} v_d \mu  -  v_u ( \kappa v_S
 + \sqrt{2} M_S )\Big){\lambda}_{H} - \frac{1}{\sqrt{2}}v_u T_{{\lambda}_{H}},\nonumber \\&&
m_{\phi_{u}{\phi}_{s}} =  \Big( {\lambda}_{H} v_u v_S  + \sqrt{2} v_u \mu
-v_d (\kappa v_S  + \sqrt{2} M_S )\Big){\lambda}_{H}
- \frac{1}{\sqrt{2}}  v_dT_{{\lambda}_{H}},\nonumber\\&&
m_{\phi_{\eta}{\phi}_{s}} = \Big( {\lambda}_{C} v_{\eta} v_S  + v_{\bar{\eta}} (\kappa v_S
 + \sqrt{2} M_S )\Big){\lambda}_{C}  +\frac{1}{\sqrt{2}}v_{\bar{\eta}} T_{{\lambda}_{C}},\nonumber\\&&
m_{\phi_{\bar{\eta}}{\phi}_{s}} =\Big( {\lambda}_{C} v_{\bar{\eta}} v_S  + v_{\eta}(\kappa v_S
 + \sqrt{2} M_S )\Big){\lambda}_{C} + \frac{1}{\sqrt{2}}v_{\eta}
 T_{{\lambda}_{C}},\nonumber\\&&
m_{{\phi}_{s}{\phi}_{s}} = m^2_{S}+ \Big(2 l_W  + 3v_S (\kappa v_S  + 2\sqrt{2} M_S )
+ {\lambda}_{C} v_{\eta} v_{\bar{\eta}}  - {\lambda}_{H} v_d v_u \Big)\kappa\nonumber \\
 && +\frac{1}{2}|{\lambda}_{C}|^2 \xi^2+\frac{1}{2}|{\lambda}_{H}|^2 v^{2}
  +2 {B_{S}}  + 4 |M_S|^2   + \sqrt{2} v_S T_{\kappa}.
\end{eqnarray}}
\begin{equation}
m^2_{A^0} = \left(
\begin{array}{ccccc}
m_{\sigma_{d}\sigma_{d}} &m_{\sigma_{u}\sigma_{d}} &m_{\sigma_{\eta}\sigma_{d}} &m_{\sigma_{\bar{\eta}}\sigma_{d}} &m_{{\sigma}_{s}\sigma_{d}}\\
m_{\sigma_{d}\sigma_{u}} &m_{\sigma_{u}\sigma_{u}} &m_{\sigma_{\eta}\sigma_{u}} &m_{\sigma_{\bar{\eta}}\sigma_{u}} &m_{{\sigma}_{s}\sigma_{u}}\\
m_{\sigma_{d}\sigma_{\eta}} &m_{\sigma_{u}\sigma_{\eta}} &m_{\sigma_{\eta}\sigma_{\eta}} &m_{\sigma_{\bar{\eta}}\sigma_{\eta}} &m_{{\sigma}_{s}\sigma_{\eta}}\\
m_{\sigma_{d}\sigma_{\bar{\eta}}} &m_{\sigma_{u}\sigma_{\bar{\eta}}} &m_{\sigma_{\eta}\sigma_{\bar{\eta}}} &m_{\sigma_{\bar{\eta}}\sigma_{\bar{\eta}}} &m_{{\sigma}_{s}\sigma_{\bar{\eta}}}\\
m_{\sigma_{d}{\sigma}_{s}} &m_{\sigma_{u}{\sigma}_{s}} &m_{\sigma_{\eta}{\sigma}_{s}} &m_{\sigma_{\bar{\eta}}{\sigma}_{s}} &m_{{\sigma}_{s}{\sigma}_{s}}\end{array}
\right).\label{CPoddhiggs}
 \end{equation}
Eq.(\ref{CPoddhiggs}) is the CP-odd Higgs mass squared matrix, whose elements are
{\small\begin{eqnarray}
&&m_{\sigma_{d}\sigma_{d}} = m_{H_d}^2+  |\mu|^2+\frac{1}{8} \Big( [g_{1}^{2}+(g_{X}+g_{YX})^{2}+g_2^2] ( v_{d}^{2}  - v_{u}^{2})
 \nonumber \\ &&+2 (g_{Y X} g_{X}+g_X^2) ( v_{\eta}^{2}- v_{\bar{\eta}}^{2})\Big)
+  \sqrt{2} v_S \mu {\lambda}_{H}  + \frac{1}{2}(v_{u}^{2} + v_S^{2})|{\lambda}_{H}|^2 ,\nonumber\\&&
m_{\sigma_{d}\sigma_{u}} =  \Big((\sqrt{2} M_S v_S  + l_W) + \frac{1}{2}\kappa v_S^{2}
+ \frac{1}{2}{\lambda}_{C} v_{\eta} v_{\bar{\eta}} \Big){\lambda}_{H}  +  B_{\mu}
+ \frac{1}{\sqrt{2}} v_S T_{{\lambda}_{H}}, \nonumber\\
&&m_{\sigma_{u}\sigma_{u}} = m_{H_u}^2+ |\mu|^2 +\frac{1}{8} \Big( [g_{1}^{2}+(g_{X}+g_{YX})^{2}+g_2^2] ( v_{u}^{2}  - v_{d}^{2})
 \nonumber \\ &&+2 (g_{Y X} g_{X}+g_X^2) ( v_{\bar{\eta}}^{2}- v_{\eta}^{2})\Big)
 +  \sqrt{2} v_S \mu {\lambda}_{H}  + \frac{1}{2}(v_{d}^{2} + v_S^{2})|{\lambda}_{H}|^2, \nonumber\\
&&m_{\sigma_{\eta}\sigma_{\eta}} = m_{\eta}^2+\frac{1}{4} \Big((g_{Y X} g_{X}+g_{X}^2)
 ( v_{d}^{2}- v_{u}^{2})+2g_{X}^{2} (v_{\eta}^{2}- v_{\bar{\eta}}^{2})\Big)+\frac{1}{2} (v_{\bar{\eta}}^{2} + v_S^{2})|{\lambda}_{C}|^2,\nonumber \\&&
m_{\sigma_{\eta}\sigma_{\bar{\eta}}} = \frac{1}{2} \Big((-2 l_W  + {\lambda}_{H} v_d v_u ){\lambda}_{C}
 - \sqrt{2} v_S (2 M_S {\lambda}_{C}   + T_{{\lambda}_{C}}) - v_S^{2} {\lambda}_{C} \kappa \Big),\nonumber
\\&&m_{\sigma_{\bar{\eta}}\sigma_{\bar{\eta}}} = m_{\bar{\eta}}^2
 +\frac{1}{4} \Big((g_{Y X} g_{X}+g_{X}^2)
 ( v_{u}^{2}- v_{d}^{2})+2g_{X}^{2} (v_{\bar{\eta}}^{2}- v_{\eta}^{2})\Big)+\frac{1}{2} (v_{\eta}^{2} + v_S^{2})|{\lambda}_{C}|^2, \nonumber\\
&& m_{\sigma_{d}{\sigma}_{s}} = - v_u \Big((\kappa v_S  + \sqrt{2} M_S )
{\lambda}_{H}  -\frac{1}{\sqrt{2}}T_{{\lambda}_{H}} \Big),~~~~~~m_{\sigma_{d}\sigma_{\eta}} = -\frac{1}{2} v_u v_{\bar{\eta}} {\lambda}_{H} {\lambda}_{C}, \nonumber\\&&
m_{\sigma_{u}{\sigma}_{s}} = - v_d \Big((\kappa v_S  + \sqrt{2} M_S )
{\lambda}_{H}  -\frac{1}{\sqrt{2}}T_{{\lambda}_{H}} \Big),~~~~~~m_{\sigma_{u}\sigma_{\eta}} = -\frac{1}{2} v_d v_{\bar{\eta}} {\lambda}_{H} {\lambda}_{C},\nonumber\\&&
m_{\sigma_{\eta}{\sigma}_{s}} =  v_{\bar{\eta}} \Big((\kappa v_S  +
\sqrt{2} M_S ){\lambda}_{C}  - \frac{1}{\sqrt{2}}T_{{\lambda}_{C}}\Big),~~~~~~~~~m_{\sigma_{d}\sigma_{\bar{\eta}}} = -\frac{1}{2} v_u v_{\eta} {\lambda}_{H} {\lambda}_{C},\nonumber\\&&
m_{\sigma_{\bar{\eta}}{\sigma}_{s}} = v_{\eta} \Big((\kappa v_S  +
\sqrt{2} M_S ){\lambda}_{C}  - \frac{1}{\sqrt{2}}T_{{\lambda}_{C}}\Big),~~~~~~~~~m_{\sigma_{u}\sigma_{\bar{\eta}}} = -\frac{1}{2} v_d v_{\eta} {\lambda}_{H} {\lambda}_{C},\nonumber\\&&
m_{{\sigma}_{s}{\sigma}_{s}} = m^2_{S}+ 4 |M_S|^2 +(\kappa v_S^{2}  -2 l_W  - {\lambda}_{C} v_{\eta} v_{\bar{\eta}}
+ {\lambda}_{H} v_d v_u )\kappa -2B_{S}\nonumber\\&&+\frac{1}{2}|{\lambda}_{C}|^2 \xi^2
 + \frac{1}{2}|{\lambda}_{H}|^2 v^{2} +\sqrt{2} v_S (2 M_S \kappa-T_{\kappa} ) .
\end{eqnarray}}
The mass matrix for slepton with the basis $(\tilde{e}_L, \tilde{e}_R)$ is diagonalized by $Z^E$ through the
formula $Z^E m^2_{\tilde{e}} Z^{E,\dagger} = m^{diag}_{2,\tilde{e}}$,
\begin{equation}
m^2_{\tilde{e}} = \left(
\begin{array}{cc}
m_{\tilde{e}_L\tilde{e}_L^*} &\frac{1}{2} \Big(\sqrt{2} v_d T_{e}^{\dagger}  - v_u \Big({\lambda}_{H} xS  + \sqrt{2} \mu \Big)Y_{e}^{\dagger} \Big)\\
\frac{1}{2} \Big(\sqrt{2} v_d T_e  - v_u Y_e \Big(\sqrt{2} \mu^*  + xS {\lambda}_{H}^* \Big)\Big) &m_{\tilde{e}_R\tilde{e}_R^*}\end{array}
\right).
 \end{equation}
\begin{eqnarray}
&&m_{\tilde{e}_L\tilde{e}_L^*} = m_{\tilde{l}}^2+\frac{1}{8} \Big((g_{1}^{2} + g_{Y X}^{2}
+ g_{Y X} g_{X} -g_2^2)(v_{d}^{2}- v_{u}^{2})+ 2 g_{Y X} g_{X}( v_{\eta}^{2}- v_{\bar{\eta}}^{2}
)
\Big)+\frac{1}{2} v_{d}^{2} {Y_{e}^{\dagger}  Y_e} ,\nonumber\\&&
m_{\tilde{e}_R\tilde{e}_R^*} = m_e^2-\frac{1}{8}  \Big([2(g_{1}^{2} + g_{Y X}^{2})+3g_{Y X} g_{X}+g_{X}^{2}]
( v_{d}^{2}- v_{u}^{2})\nonumber\\&&\hspace{2.0cm}+(4g_{Y X} g_{X}+2g_{X}^{2})(v_{\eta}^{2}- v_{\bar{\eta}}^{2})
\Big)+\frac{1}{2} v_{d}^{2} {Y_e  Y_{e}^{\dagger}}.
\end{eqnarray}

The mass matrix for neutralino in the basis $(\lambda_{\tilde{B}}, \tilde{W}^0, \tilde{H}_d^0, \tilde{H}_u^0,
\lambda_{\tilde{X}}, \tilde{\eta}, \tilde{\bar{\eta}}, \tilde{s}) $ is,

\begin{equation}
m_{\tilde{\chi}^0} = \left(
\begin{array}{cccccccc}
M_1 &0 &-\frac{g_1}{2}v_d &\frac{g_1}{2}v_u &{M}_{B B'} &0  &0  &0\\
0 &M_2 &\frac{1}{2} g_2 v_d  &-\frac{1}{2} g_2 v_u  &0 &0 &0 &0\\
-\frac{g_1}{2}v_d &\frac{1}{2} g_2 v_d  &0
&m_{\tilde{H}_u^0\tilde{H}_d^0} &m_{\lambda_{\tilde{X}}\tilde{H}_d^0} &0 &0 & - \frac{{\lambda}_{H} v_u}{\sqrt{2}}\\
\frac{g_1}{2}v_u &-\frac{1}{2} g_2 v_u  &m_{\tilde{H}_d^0\tilde{H}_u^0} &0 &m_{\lambda_{\tilde{X}}\tilde{H}_u^0} &0 &0 &- \frac{{\lambda}_{H} v_d}{\sqrt{2}}\\
{M}_{B B'} &0 &m_{\tilde{H}_d^0\lambda_{\tilde{X}}} &m_{\tilde{H}_u^0\lambda_{\tilde{X}}} &{M}_{BL} &- g_{X} v_{\eta}  &g_{X} v_{\bar{\eta}}  &0\\
0  &0 &0 &0 &- g_{X} v_{\eta}  &0 &\frac{1}{\sqrt{2}} {\lambda}_{C} v_S  &\frac{1}{\sqrt{2}} {\lambda}_{C} v_{\bar{\eta}} \\
0  &0 &0 &0 &g_{X} v_{\bar{\eta}}  &\frac{1}{\sqrt{2}} {\lambda}_{C} v_S  &0 &\frac{1}{\sqrt{2}} {\lambda}_{C} v_{\eta} \\
0 &0 & - \frac{{\lambda}_{H} v_u}{\sqrt{2}} &- \frac{{\lambda}_{H} v_d}{\sqrt{2}} &0 &\frac{1}{\sqrt{2}} {\lambda}_{C} v_{\bar{\eta}}
 &\frac{1}{\sqrt{2}} {\lambda}_{C} v_{\eta}  &m_{\tilde{s}\tilde{s}}\end{array}
\right),\label{neutralino}
 \end{equation}

\begin{eqnarray}
&& m_{\tilde{H}_d^0\tilde{H}_u^0} = - \frac{1}{\sqrt{2}} {\lambda}_{H} v_S  - \mu ,~~~~~~~
m_{\tilde{H}_d^0\lambda_{\tilde{X}}} = -\frac{1}{2} \Big(g_{Y X} + g_{X}\Big)v_d, \nonumber\\&&
m_{\tilde{H}_u^0\lambda_{\tilde{X}}} = \frac{1}{2} \Big(g_{Y X} + g_{X}\Big)v_u
 ,~~~~~~~~~~~~
m_{\tilde{s}\tilde{s}} = 2 M_S  + \sqrt{2} \kappa v_S.\label{neutralino1}
\end{eqnarray}
This matrix is diagonalized by $Z^N$
\begin{equation}
Z^{N*} m_{\tilde{\chi}^0} Z^{N{\dagger}} = m^{diag}_{\tilde{\chi}^0}.
\end{equation}

Here, we show the needed couplings in this model.
The CP-even Higgs couple with CP-even sneutrinos
\begin{eqnarray}
&&\mathcal{L}_{H\tilde{\nu}^R\tilde{\nu}^R}=H_i\tilde{\nu}^R_j\frac{i}{4}\Big\{\sum_{a,b=1}^3\Big[-2\sqrt{2}Z^{R*}_{kb}Z^{R*}_{j 3+a}(T_\nu)_{ab}Z^{H}_{i2}-2\lambda_C v_S Z^{R*}_{k3+b}Z^{R*}_{j 3+a}(Y_X)_{ab}Z^{H}_{i3}\nonumber\\&&
-2\sqrt{2}Z^{R*}_{k3+b}Z^{R*}_{j 3+a}(T_X)_{ab}Z^{H}_{i4}-2\lambda_C v_\eta Z^{R*}_{k3+b}Z^{R*}_{j 3+a}(Y_X)_{ab}Z^{H}_{i5}
\Big]+\Big[j\leftrightarrow k\Big]\nonumber\\&&
-16v_{\bar{\eta}}\sum_{a,b,c=1}^3 Z^{R*}_{k3+c}Z^{R*}_{j 3+b}(Y_X)_{ac}(Y_X)_{ab}Z^{H}_{i4}+
\sum_{a=1}^3Z^{R*}_{ka}Z^{R*}_{ja}\Big[(g_{YX}g_X+g_1^2\nonumber\\&&+g_{YX}^2+g_2^2)(-v_dZ^H_{i1}+v_uZ^H_{i2})-2g_{YX}g_X(-v_{\bar{\eta}}Z^H_{i4}+v_\eta Z^H_{i3})\Big]
\nonumber\\&&+
\sum_{a=1}^3Z^{R*}_{k3+a}Z^{R*}_{j3+a}\Big[(g_{YX}g_X+g_{X}^2)(v_uZ^H_{i2}-v_dZ^H_{i1})-2g_X^2(v_\eta Z^H_{i3}-v_{\bar{\eta}}Z^H_{i4})\Big]
\Big\}\tilde{\nu}^{*R}_k.
\end{eqnarray}
The coupling of two CP-even Higgs and two CP-even sneutrinos reads as
\begin{eqnarray}
&&\mathcal{L}_{HH\tilde{\nu}^R\tilde{\nu}^R}=H_i\tilde{\nu}^R_l\Big\{\frac{i}{2}\sum_{a,b=1}^3\Big[\Big(-\lambda_CZ^{R*}_{l3+b}
Z^{R*}_{k3+a}(Y_X)_{ab}(Z^H_{i5}Z^H_{j3}+Z^H_{i3}Z^H_{j5})\Big)+\Big(l\leftrightarrow k\Big)
\Big]\nonumber\\&&+\frac{i}{4}\sum_{a=1}^3Z^{R*}_{l3+a}Z^{R*}_{k3+a}\Big[(g_{YX}g_X+g_{X}^2)
(Z^H_{i2}Z^H_{j2}-Z^H_{i1}Z^H_{j1})-2g_X^2(Z^H_{i3}Z^H_{j3}-Z^H_{i4}Z^H_{j4})
\Big]\nonumber\\&&
+\frac{i}{4}\sum_{a=1}^3Z^{R*}_{la}Z^{R*}_{ka}\Big[(g_{YX}g_X+g_1^2+g_{YX}^2+g_2^2)(-Z^H_{i1}Z^H_{j1}+Z^H_{i2}Z^H_{j2})
\nonumber\\&&-2g_{YX}g_X(Z^H_{i3}Z^H_{j3}-Z^H_{i4}Z^H_{j4})
\Big]-4i\sum_{a,b,c=1}^3Z^{R*}_{l3+c}Z^{R*}_{k3+b}(Y_X)_{ab}(Y_X)_{ac}Z^H_{i4}Z^H_{j4}
\Big\}H_k\tilde{\nu}^R_k .
\end{eqnarray}
 The other used vertexes including the couplings of:
$H-H-H,~ H-W-W$ and $H-Z-Z$ are
{\small\begin{eqnarray}
&&\mathcal{L}_{HHH}=iH_iH_j\Big\{(\frac{1}{4}g_1^2+\frac{1}{4}g_{YX}^2+\frac{1}{4}g_2^2+\frac{1}{2}g_{YX}g_X+\frac{1}{4}g_X^2
-\lambda_H^2)[v_u\langle112\rangle+v_d\langle122\rangle]\nonumber\\&&
-(\frac{3}{4}g_1^2+\frac{3}{4}g_{YX}^2+\frac{3}{4}g_2^2+\frac{3}{2}g_{YX}g_X+\frac{3}{4}g_X^2)[v_u\langle111\rangle+v_d\langle222\rangle]
+\frac{1}{2}(g_{YX}g_X+g_X^2)\nonumber\\&&\times\Big[v_{\bar{\eta}}(\langle114\rangle+\langle224\rangle)-v_\eta(\langle113\rangle+\langle223\rangle)
+v_u(\langle233\rangle+\langle244\rangle)-v_d(\langle133\rangle+\langle144\rangle)
\Big]\nonumber\\&&-(v_S\lambda_H^2+\sqrt{2}\mu\lambda_H)(\langle115\rangle+\langle225\rangle)+
(\lambda_Hv_S\kappa+\sqrt{2}M_S\lambda_H+\frac{1}{\sqrt{2}}T_{\lambda_H})\langle125\rangle\nonumber\\&&
-(\lambda_Cv_S\kappa+\sqrt{2}M_S\lambda_C+\frac{1}{\sqrt{2}}T_{\lambda_C})\langle345\rangle
+\frac{1}{2}\lambda_H\lambda_C\Big[v_{\bar{\eta}}\langle123\rangle+v_{\eta}\langle124\rangle+v_u\langle134\rangle\nonumber\\&&+v_d\langle234\rangle\Big]
+(\lambda_Hv_u\kappa-v_d\lambda_H^2 )\langle155\rangle+(\lambda_Hv_d\kappa-v_u\lambda_H^2 )\langle255\rangle
-3g_X^2(v_\eta\langle333\rangle+v_{\bar{\eta}}\langle444\rangle)\nonumber\\&&+(g_X^2-\lambda_C^2)(v_\eta\langle344\rangle+v_{\bar{\eta}}\langle334\rangle)
-v_S\lambda_C^2(\langle335\rangle+\langle445\rangle)-(\lambda_C^2v_\eta+\lambda_Cv_{\bar{\eta}}\kappa)\langle355\rangle\nonumber\\&&
-(\lambda_C^2v_{\bar{\eta}}+\lambda_Cv_{\eta}\kappa)\langle455\rangle-(6v_S\kappa^2+6\sqrt{2}M_S\kappa+\sqrt{2}T_{\kappa})\langle555\rangle
\Big\}H_k,\nonumber\\&&
 \mathcal{L}_{HWW}=H_iW_\mu\Big(\frac{i}{2}g_2^2(v_dZ^H_{i1}+v_uZ^H_{i2})g^{\sigma\mu}\Big)W^*_\sigma,\nonumber\\&&
  \mathcal{L}_{HZZ}=H_iZ_\mu\Big\{\frac{i}{2}\Big[\Big(g_1\cos\theta'_W\sin\theta_W+g_2\cos\theta'_W\cos\theta_W-g_{YX}g_X\sin\theta'_W\Big)^2
  \nonumber\\&&\hspace{1.5cm}\times(v_dZ^H_{i1}+v_uZ^H_{i2})
+4(g_X\sin\theta'_W)^2(v_{\bar{\eta}}Z^H_{i4}+v_\eta Z^H_{i3})
\Big]g^{\sigma\mu}\Big\}Z^*_\sigma.
\end{eqnarray}}
Here $\langle\alpha\alpha\alpha\rangle, \langle\alpha\alpha\beta\rangle, \langle\alpha\beta\gamma\rangle$ are the shorthand notations
{\small\begin{eqnarray}&&
 \langle\alpha\alpha\alpha\rangle=Z^H_{i\alpha}Z^H_{j\alpha}Z^H_{k\alpha},
~~~\langle\alpha\alpha\beta\rangle=Z^H_{i\alpha}Z^H_{j\alpha}Z^H_{k\beta}
+Z^H_{i\alpha}Z^H_{j\beta}Z^H_{k\alpha}+Z^H_{i\beta}Z^H_{j\alpha}Z^H_{k\alpha},(\alpha\neq\beta),\nonumber\\&&
\langle\alpha\beta\gamma\rangle=Z^H_{i\alpha}Z^H_{j\gamma}Z^H_{k\beta}+Z^H_{i\gamma}Z^H_{j\alpha}Z^H_{k\beta}
+Z^H_{i\alpha}Z^H_{j\beta}Z^H_{k\gamma}+Z^H_{i\gamma}Z^H_{j\beta}Z^H_{k\alpha}+Z^H_{i\beta}Z^H_{j\alpha}Z^H_{k\gamma}\nonumber\\&&\hspace{1.6cm}
+Z^H_{i\beta}Z^H_{j\gamma}Z^H_{k\alpha},~~~~(\alpha\neq\beta\neq\gamma).
\end{eqnarray}}
Some other used couplings are shown as
{\small\begin{eqnarray}
&&\mathcal{L}_{WW\tilde{\nu}^R\tilde{\nu}^R}=\tilde{\nu}^R_iW_\nu\Big(\frac{i}{2}g_2^2\sum_{a=1}^3Z^{R*}_{ia}Z^{R*}_{ja}g^{\mu\nu}\Big)\tilde{\nu}^R_jW_\mu,
\nonumber\\
&&\mathcal{L}_{ZZ\tilde{\nu}^R\tilde{\nu}^R}=\tilde{\nu}^R_iZ_\nu\Big\{i\sum_{a=1}^3\Big[Z^{R*}_{ia}Z^{R*}_{ja}
\Big(\frac{1}{2}g_2^2(\cos\theta_W\cos\theta'_W)^2+\frac{1}{2}g_1^2(\sin\theta_W\cos\theta'_W)^2\nonumber\\&&+g_1g_2\cos\theta_W\sin\theta_W(\cos\theta'_W)^2
-g_{YX}\sin\theta'_W\cos\theta'_W (g_2\cos\theta_W+g_1\sin\theta_W)\nonumber\\&&+
\frac{1}{2}g_{YX}^2(\sin\theta'_W)^2\Big)+\frac{1}{2}g_{X}^2(\sin\theta'_W)^2Z^{R*}_{i3+a}Z^{R*}_{j3+a}
\Big]g^{\mu\nu}\Big\}\tilde{\nu}^R_jZ_\mu,\nonumber\\&&
\mathcal{L}_{\tilde{e}\tilde{\nu}^{R*}W}=\tilde{e}_i\tilde{\nu}^{R*}_j\Big(-\frac{i}{2}g_2\sum_{a=1}^3Z^{E*}_{ia}Z^{R*}_{ja}(-p_{\mu}^{\tilde{\nu}_j^R}+p_\mu^{\tilde{e}_i})\Big)W^\mu
+h.c,\nonumber\\&&
\mathcal{L}_{Zdd}=\bar{d}\Big[\Big(\frac{i}{6}(3g_2\cos\theta_W\cos\theta'_W+g_1\sin\theta_W\cos\theta'_W-g_{YX}\sin\theta'_W)\gamma_\mu P_L\nonumber\\&&
-\frac{i}{6}(2g_1\sin\theta_W\cos\theta'_W-(2g_{YX}+3g_X)\sin\theta'_W)\gamma_\mu P_R\Big]d~Z^\mu,
\nonumber\\&&\mathcal{L}_{Z^\prime dd}=\bar{d}\Big[-\frac{i}{6}(3g_2\cos\theta_W\sin\theta'_W+g_1\sin\theta_W\sin\theta'_W+g_{YX}\cos\theta'_W)\gamma_\mu P_L\nonumber\\&&
+\frac{i}{6}[2g_1\sin\theta_W\sin\theta'_W+(2g_{YX}+3g_X)\cos\theta'_W]\gamma_\mu P_R\Big]d~Z^{\prime\mu},\nonumber\\&&
\mathcal{L}_{Z ll}=
\bar{l}\Big\{\frac{i}{2}(-g_1\sin\theta_W\cos\theta'_W +g_2\cos\theta_W\cos\theta'_W+g_{YX}\sin\theta'_W)\gamma_\mu P_L\nonumber\\&&
-\frac{i}{2}(2g_1\sin\theta_W\cos\theta'_W-(2g_{YX}+g_X)\sin\theta'_W)\gamma_\mu P_R\Big\}lZ^\mu,
\nonumber\\
&&\mathcal{L}_{Z^{\prime} ll}=\bar{l}\Big\{\frac{i}{2}(g_1\sin\theta_W\sin\theta'_W -g_2\cos\theta_W\sin\theta'_W+g_{YX}\cos\theta'_W)\gamma_\mu P_L\nonumber\\&&
+\frac{i}{2}(2g_1\sin\theta_W\sin\theta'_W+(2g_{YX}+g_X)\cos\theta'_W)\gamma_\mu P_R\Big\}lZ^{\prime \mu},
\nonumber\\
&&\mathcal{L}_{Z uu}=\bar{u}\Big\{-\frac{i}{6}(3g_2\cos\theta_W\cos\theta'_W-g_1\sin\theta_W\cos\theta'_W+g_{YX}\sin\theta'_W)\gamma_\mu P_L\nonumber\\&&
+\frac{i}{6}[-(4g_{YX}+3g_X)\sin\theta'_W+4g_1\sin\theta_W\cos\theta'_W]\gamma_\mu P_R\Big\}uZ^\mu,
\nonumber\\
&&\mathcal{L}_{Z^{\prime} uu}=\bar{u}\Big\{-\frac{i}{6}(-3g_2\cos\theta_W\sin\theta'_W+g_1\sin\theta_W\sin\theta'_W+g_{YX}\cos\theta'_W)\gamma_\mu P_L\nonumber\\&&
-\frac{i}{6}[(4g_{YX}+3g_X)\cos\theta'_W+4g_1\sin\theta_W\sin\theta'_W]\gamma_\mu P_R\Big\}uZ^{\prime\mu},
\nonumber\\
&&\mathcal{L}_{Z\nu\nu}=\bar{\nu}_i\Big\{-\frac{i}{2}(g_1\sin\theta_W\cos\theta'_W+g_2\cos\theta_W\cos\theta'_W-g_{YX}\sin\theta'_W)\sum_{a=1}^3U^{V*}_{ja}U^{V}_{ia}\gamma_\mu P_L\nonumber\\&&
+\frac{i}{2}(g_1\sin\theta_W\cos\theta'_W+g_2\cos\theta_W\cos\theta'_W-g_{YX}\sin\theta'_W)\sum_{a=1}^3U^{V}_{ja}U^{V*}_{ia}\gamma_\mu P_R\Big\}\nu_jZ^\mu,
\nonumber\\
&&\mathcal{L}_{Z^{\prime}\nu\nu}=\bar{\nu}_i\Big\{\frac{i}{2}(g_1\sin\theta_W\sin\theta'_W+g_2\cos\theta_W\sin\theta'_W+g_{YX}\cos\theta'_W)\sum_{a=1}^3U^{V*}_{ja}U^{V}_{ia}\gamma_\mu P_L\nonumber\\&&
-\frac{i}{2}(g_1\sin\theta_W\sin\theta'_W+g_2\cos\theta_W\sin\theta'_W+g_{YX}\cos\theta'_W)\sum_{a=1}^3U^{V*}_{ia}U^{V}_{ja}\gamma_\mu P_R\Big\}\nu_jZ^{\prime\mu},
\nonumber\\
&&\mathcal{L}_{\tilde{\nu}^I\tilde{\nu}^RZ}=\tilde{\nu}^I_i\tilde{\nu}^R_j\Big\{\frac{1}{2}(-p_{\mu}^{\tilde{\nu}^R_j}+p_{\mu}^{\tilde{\nu}^I_i})\Big[
\Big(g_1\sin\theta_W\cos\theta'_W+g_2\cos\theta_W\cos\theta'_W\nonumber\\&&-g_{YX}\sin\theta'_W\Big)\sum_{a=1}^3Z^{I*}_{ia}Z^{R*}_{ja}
+g_X\sin\theta'_W\sum_{a=1}^3Z^{I*}_{i3+a}Z^{R*}_{j3+a}\Big]\Big\}Z^\mu,
\nonumber\\
&&\mathcal{L}_{\tilde{\nu}^I\tilde{\nu}^RZ}=\tilde{\nu}^I_i\tilde{\nu}^R_j\Big\{\frac{1}{2}(-p_{\mu}^{\tilde{\nu}^R_j}+p_{\mu}^{\tilde{\nu}^I_i})\Big[
-\Big(g_1\sin\theta_W\sin\theta'_W+g_2\cos\theta_W\sin\theta'_W\nonumber\\&&+g_{YX}\cos\theta'_W\Big)\sum_{a=1}^3Z^{I*}_{ia}Z^{R*}_{ja}
+g_X\cos\theta'_W\sum_{a=1}^3Z^{I*}_{i3+a}Z^{R*}_{j3+a}\Big]\Big\}Z^{\prime\mu}.
\end{eqnarray}}


\begin{thebibliography}{50}

\vspace{3mm}

\bibitem{account1}P.A.R. Ade, et al., \emph{Planck 2013 results. XVI. Cosmological parameters
Planck Collaboration}, Astron. Astrophys. {\bf571} (2014) A16 [arXiv: 1303. 5076].
\bibitem{account2}
R.H. Cyburt, \emph{Primordial nucleosynthesis for the new cosmology: Determining uncertainties and examining concordance},
Phys. Rev. D {\bf70} (2004) 023505 [astro-ph/0401091].
\bibitem{rotation1}G. Bertone, D. Hooper, J. Silk, \emph{Particle dark matter: Evidence, candidates and constraints}, Phys. Rept. {\bf405} (2005) 279
[hep-ph/0404175].
\bibitem{rotation2}E. Corbelli, P. Salucci, \emph{The Extended Rotation Curve and the Dark Matter Halo of M33},
 Mon. Not. Roy. Astron. {\bf311} (2000) 441 [astro-ph/9909252].

\bibitem{other exist1}D. Clowe, M. Bradac, A.H. Gonzalez, et al.,
\emph{A direct empirical proof of the existence of dark matter}, Astrophys. J. {\bf648} (2006) 109 [astro-ph/0608407].
\bibitem{other exist2}A. Taylor, S. Dye, T.J. Broadhurst, et al., \emph{Gravitational lens magnification
and the mass of abell 1689}, Astrophys. J. {\bf501} (1998) 539 [astro-ph/9801158].
\bibitem{other exist3}D. Walsh, R.F. Carswell, R.J. Weymann.\emph{0957 + 561 A, B - Twin quasistellar objects or gravitational lens}, Nature {\bf279} (1979) 381.
\bibitem{other exist4}J.L. Feng, \emph{Dark Matter Candidates from Particle Physics and Methods of Detection}, Ann. Rev. Astron. Astrophys. {\bf48} (2010) 495 [arXiv: 1003. 0904].

\bibitem{longlife1}M. Drees, M.M. Nojiri, \emph{The Neutralino relic density in minimal N=1 supergravity}, Phys. Rev. D {\bf47} (1993) 376 [hep-ph/9207234].

\bibitem{longlife2} L.B. Jia, \emph{Dark photon portal dark matter with the 21-cm anomaly}, Eur. Phys. J. C {\bf79} (2019) 80 [arXiv: 1804. 07934].

\bibitem{mh01}CMS Collaboration, \emph{Observation of a New Boson at a Mass of 125 GeV with the CMS Experiment at the LHC},
Phys. Lett. B {\bf 716} (2012) 30 [arXiv: 1207. 7235].
\bibitem{mh02} ATLAS Collaboration, \emph{Observation of a new particle in the search for the Standard Model Higgs
boson with the ATLAS detector at the LHC}, Phys. Lett. B {\bf 716} (2012) 1 [arXiv: 1207. 7214].

\bibitem{pdg}Particle Data Group collaboration, \emph{Review of Particle Physics}, Phys. Rev. D {\bf98} (2018) 030001.

\bibitem{WIMP}S. Andreas, T. Hambye,\emph{ WIMP dark matter, Higgs exchange and DAMA}, JCAP {\bf0810} (2008) 034 [arXiv: 0808. 0255].
\bibitem{WIMP1}J.J. Cao, Z.X. Heng, J.M. Yang, et al., \emph{Higgs decay to dark matter in low energy SUSY: is it detectable at the LHC ?} JHEP {\bf1206}
 (2012) 145 [arXiv: 1203. 0694].

\bibitem{MSSM}J. Rosiek, \emph{Complete set of Feynman rules for the MSSM: Erratum}, Phys. Rev. D {\bf41} (1990) 3464 [Erratum: hep-ph/9511250].

\bibitem{LSneu}Z. Thomas, D.T. Smith, N. Weiner, \emph{Mixed Sneutrinos, Dark Matter and the CERN LHC}, Phys. Rev. D {\bf77} (2008) 115015 [arXiv: 0712. 4146].

\bibitem{neutrino1} T2K Collab, \emph{Indication of Electron Neutrino Appearance from an
Accelerator-produced Off-axis Muon Neutrino Beam}, Phys. Rev. Lett. {\bf 107} (2011) 041801 [arXiv: 1106. 2822].
 \bibitem{neutrino2}
MINOS Collab, \emph{Improved search for muon-neutrino to electron-neutrino oscillations in MINOS}, Phys. Rev. Lett. {\bf 107} (2011) 181802 [arXiv: 1108. 0015].


\bibitem{Sneudark11}A. Ghosh, T. Mondal, B. Mukhopadhyaya, \emph{Right sneutrino with ¦¤L=2 masses
as nonthermal dark matter}, Phys. Rev. D {\bf99} (2019) 035018 [arXiv: 1807. 04964].

\bibitem{Sneudark12}C. Arina, N. Fornengo, \emph{Sneutrino cold dark matter, a new analysis: Relic abundance
and detection rates}, JHEP {\bf0711} (2007) 029 [arXiv: 0709. 4477].

\bibitem{Sneudark13}C. Arina, F. Bazzocchi, N. Fornengo, et al,\emph{ Minimal supergravity sneutrino dark matter and inverse seesaw neutrino masses},
 Phys. Rev. Lett. {\bf 101} (2008) 161802 [arXiv: 0806. 3225].

\bibitem{Sneudark14}H.N. Long, \emph{Right-handed sneutrinos as self-interacting dark matter in supersymmetric
economical 3-3-1 model}, Adv. Stud. Theor. Phys. {\bf 4} (2010) 173 [arXiv: 0710. 5833].

\bibitem{Sneudark15}D.G. Cerdeno, O. Seto, \emph{Right-handed sneutrino dark matter in the NMSSM}, JCAP {\bf 0908} (2009) 032 [arXiv: 0903. 4677].

\bibitem{Caojunjie1}
J.J. Cao, X.F. Guo, Y.L. He, et al., \emph{Sneutrino DM in the NMSSM with inverse seesaw mechanism}, JHEP {\bf 1710} (2017) 044
[arXiv: 1707. 09626].
\bibitem{Caojunjie2}	
J.J. Cao, X.F. Guo, Y.S. Pan, et al., \emph{Bayesian analysis of sneutrino dark matter in the NMSSM with a type-I seesaw mechanism},
Phys. Rev. D {\bf99} (2019) 115033 [arXiv: 1807. 03762].

\bibitem{TaoHan}
T. Han, H.K. Liu, S. Mukhopadhyay, et al.,\emph{ Dark Matter Blind Spots at One-Loop}, JHEP {\bf 1903} (2019) 080 [arXiv: 1810. 04679].

\bibitem{Sneudark21}
J.M. Russell, C. McCabe, M. McCullough, \emph{Neutrino-Flavoured Sneutrino Dark Matter}, JHEP {\bf 1003} (2010) 108 [arXiv: 0911. 4489].

\bibitem{Sneudark22}D.A. Demir, L.L. Everett, M. Frank, et al., \emph{Sneutrino Dark Matter:
Symmetry Protection and Cosmic Ray Anomalies}, Phys. Rev. D {\bf81} (2010) 035019 [arXiv: 0906. 3540].

\bibitem{Sneudark23} Z.F. Kang, J.M. Li, T.J. Li, et al., \emph{The maximal $U(1)_L$ inverse seesaw from  d=5  operator and oscillating asymmetric Sneutrino dark matter}, Eur. Phys. J. C {\bf76} (2016) 270 [arXiv: 1102. 5644].

\bibitem{Sneudark24}D.G. Cerdeno, J.H. Huh, M. Peiro, et al., \emph{Very light right-handed sneutrino dark matter in the NMSSM}, JCAP {\bf1111} (2011) 027 [arXiv: 1108. 0978].

\bibitem{sneutrinoD1}
B. Zhu, R. Ding, Y. Li, \emph{Realization of Sneutrino Self-interacting Dark Matter in the Focus Point Supersymmetry},
Phys. Rev. D {\bf 98} (2018) 035007 [arXiv: 1804. 00277].
\bibitem{sneutrinoD2}
J. Chang, K.M. Cheung, H. Ishida, et al., \emph{Sneutrino Dark Matter via pseudoscalar X-funnel meets Inverse Seesaw}, JHEP {\bf 1809} (2018) 071 [arXiv: 1806. 04468].
\bibitem{sneutrinoD3}
D.K. Ghosh, K. Huitu, S. Mondal, \emph{Same-sign trilepton signal for stop quark in the presence of sneutrino dark matter}, Phys. Rev. D {\bf99} (2019) 075014 [arXiv: 1807. 07385].

\bibitem{UMSSM1}H.S. Lee, K.T. Matchev, S. Nasri, \emph{Revival of the thermal sneutrino dark matter}, Phys. Rev. D {\bf76} (2007) 041302 [hep-ph/0702223].

\bibitem{UMSSM2}P. Bandyopadhyay, E.J. Chun, J.C. Park, \emph{Right-handed sneutrino dark matter in  U(1)'  seesaw
models and its signatures at the LHC}, JHEP {\bf1106} (2011) 129
[arXiv: 1105. 1652].

\bibitem{UMSSM3}G. Belanger, J.D. Silva, A. Pukhov, \emph{The Right-handed sneutrino as
thermal dark matter in U(1) extensions of the MSSM}, JCAP {\bf1112} (2011) 014 [arXiv: 1110. 2414].

\bibitem{UMSSM4}G. Belanger, J.D. Silva, U. Laa, et al., \emph{Probing U(1) extensions of the MSSM at
the LHC Run I and in dark matter searches}, JHEP {\bf 1509} (2015) 151 [arXiv: 1505. 06243].

\bibitem{UMSSM5}G. Belanger, J.D. Silva, H.M. Tran, \emph{Dark matter in U(1)
extensions of the MSSM with gauge kinetic mixing}, Phys. Rev. D {\bf95} (2017) 115017 [arXiv: 1703. 03275].

\bibitem{Sarah1}	F. Staub, \emph{Sarah}, arXiv: 0806.0538.
\bibitem{Sarah2}	 F. Staub, \emph{SARAH 4: A tool for (not only SUSY) model builders}, Comput. Phys. Commun. {\bf185} (2014) 1773 [arXiv: 1309. 7223].
\bibitem{Sarah3}	F. Staub, \emph{Exploring new models in all detail with SARAH}, Adv. High Energy Phys. {\bf2015} (2015) 840780 [arXiv: 1503. 04200].

\bibitem{PAMELA}
 H.E.S.S. Collaboration (F. Aharonian  et al.),
 \emph{The energy spectrum of cosmic-ray electrons at TeV energies}, Phys. Rev. Lett. {\bf101} (2008) 261104 [arXiv: 0811. 3894].

\bibitem{pzhandpfu} O. Adriani et al., \emph{A new measurement of the antiproton-to-proton flux ratio up to 100 GeV in the cosmic radiation}, Phys. Rev. Lett. {\bf102} (2009) 051101
[arXiv: 0810. 4994].

\bibitem{LCTHiggs1} M. Carena, J.R. Espinosaos, C.E.M. Wagner, et al., \emph{Analytical expressions for radiatively corrected Higgs masses and couplings in the MSSM}, Phys. Lett. B {\bf355} (1995) 209 [hep-ph/9504316].
\bibitem{LCTHiggs2}M. Carena, S. Gori, N.R. Shah, et al., \emph{A 125 GeV SM-like Higgs in the MSSM and the $\gamma\gamma$ rate}, JHEP {\bf 1203} (2012) 014 [arXiv: 1112. 3336].

\bibitem{Peskin} M.E. Peskin, D.V. Schroeder, \emph{An introduction to quantum field theory}, Addison Wesley,
Reading, USA, 1995.

\bibitem{B-L1}V. Barger, P.F. Perez, S. Spinner, \emph{Minimal gauged U(1)(B-L) model with spontaneous R-parity violation}, Phys. Rev. Lett.
{\bf 102} (2009) 181802 [arXiv: 0812. 3661].
 	
\bibitem{B-L2}
P.H. Chankowski, S. Pokorski, J. Wagner, \emph{Z-prime and the Appelquist-Carrazzone decoupling}, Eur. Phys. J. C {\bf 47} (2006) 187
 [hep-ph/0601097].

\bibitem{gaugemass}J.L. Yang, T.F. Feng, S.M. Zhao, et al., \emph{Two loop electroweak corrections to $\bar{B}\rightarrow X_S\gamma$ and $B_S^0\rightarrow \mu^+\mu^-$ in the B-LSSM}, Eur. Phys. J. C {\bf78} (2018) 714 [arXiv: 1803. 09904].

\bibitem{boltzmann11}J. McDonald,  \emph{Gauge singlet scalars as cold dark matter}, Phys. Rev. D {\bf50} (1994) 3637 [hep-ph/0702143].
\bibitem{boltzmann12}G. B¨¦langer, F. Boudjema, \emph{micrOMEGAs4.1: two dark matter candidates},  Comput. Phys. Commun. {\bf192} (2015) 322 [arXiv: 1407. 6129].
\bibitem{XFBO1}G. Jungman, M. Kamionkowski, K. Griest, \emph{Supersymmetric dark matter}, Phys. Rep. {\bf 267} (1996) 195 [hep-ph/9506380].


\bibitem{importantGS} S. Gopalakrishna, A.D. Gouvea, W. Porod, \emph{Right-handed sneutrinos as nonthermal dark matter}, JCAP {\bf0605} (2006) 005 [hep-ph/0602027].

\bibitem{HXG}X.G. He, T. Li, \emph{Constraints on Scalar Dark Matter from Direct Experimental Searches}, Phys. Rev. D {\bf79} (2009) 023521 [arXiv: 0811. 0658].
\bibitem{XFCW}
W. Chao, \emph{Majorana Dark matter with B+L gauge symmetry}, JHEP {\bf 1704} (2017) 034 [arXiv: 1604. 01771].


\bibitem{zhaosm}S.M. Zhao, T.F. Feng, et al., \emph{The extended BLMSSM with a 125 GeV Higgs boson and dark matter}, Eur. Phys. J. C {\bf78} (2018) 324 [arXiv: 1711. 10731].

\bibitem{LJandHE}M. Freytsis, Z. Ligeti, \emph{On dark matter models with uniquely spin-dependent detection possibilities}, Phys. Rev. D {\bf83} (2011) 115009 [arXiv: 1012. 5317].

\bibitem{DarkSUSY1}T. Bringmann, J. Edsjo, P. Gondolo, et al., \emph{DarkSUSY 6: An Advanced Tool to Compute Dark Matter Properties Numerically}, JCAP {\bf1807} (2018) 033 [arXiv: 1802. 03399].
\bibitem{DarkSUSY2}    	
G. Belanger, F. Boudjema, A. Goudelis, et al., \emph{micrOMEGAs5.0: Freeze-in}, Comput. Phys. Commun. {\bf231} (2018) 173 [arXiv: 1801. 03509].
\bibitem{DarkSUSY3}   W. Chao, \emph{Direct detections of Majorana dark matter in vector portal}, JHEP {\bf1911} (2019) 013  [arXiv: 1904. 09785].


\bibitem{low energy1} V.D. Barger, K.M. Cheung, K. Hagiwara, et al., \emph{Global study of electron  quark contact interactions},
Phys. Rev. D {\bf 57} (1998) 391 [hep-ph/9707412].

\bibitem{low energy2} J. Erler, M.J.R. Musolf, \emph{Low energy tests of the weak interaction}, Prog. Part. Nucl. Phys. {\bf 54} (2005) 351
[hep-ph/0404291].

\bibitem{ZP1}
    CMS Collab., \emph{Search for heavy resonances decaying to tau lepton pairs in proton-proton collisions at $\sqrt{s}=13$ TeV},
     JHEP {\bf1702} (2017) 048 [arXiv: 1611. 06594].

\bibitem{ZP2} D. Pappadopulo et al., \emph{Heavy Vector Triplets: Bridging Theory and Data}, JHEP {\bf1409} (2014) 060 [arXiv: 1402. 4431].

 \bibitem{ATLAS2016} ATLAS Collaboration,\emph{ Search for new high-mass phenomena in the dilepton final state using 36 $fb^{-1}$  of proton-proton collision data at  $\sqrt{s}=13$ TeV with the ATLAS detector}, JHEP {\bf1710} (2017) 182 [arXiv:1707.02424].

\bibitem{ZPG1} G. Cacciapaglia, C. Csaki, G. Marandella, et al., \emph{The Minimal Set of Electroweak Precision Parameters},
Phys. Rev. D {\bf74} (2006) 033011 [hep-ph/0604111].

\bibitem{ZPG2} M. Carena, A. Daleo, B. A. Dobrescu, et al., \emph{Z¡ägauge bosons at the Tevatron}, Phys. Rev. D {\bf70} (2004) 093009 [hep-ph/0408098].

\bibitem{TanBP} L. Basso, \emph{The Higgs sector of the minimal SUSY B-L model}, Adv. High Energy Phys. {\bf2015} (2015) 980687 [arXiv: 1504. 05328].

\bibitem{PanXen1} PandaX-II Collaboration, \emph{Dark Matter Results From 54-Ton-Day Exposure of PandaX-II Experiment}, Phys. Rev. Lett. {\bf119} (2017) 181302 [arXiv: 1708. 06917].

\bibitem{PanXen2} XENON Collaboration, \emph{First Dark Matter Search Results from the XENON1T Experiment}, Phys. Rev. Lett. {\bf119} (2017) 181301 [arXiv: 1705. 06655].


\end{thebibliography}
\end{document}